%% file: main.tex
\documentclass[twocolumn, dvipsnames, superscriptaddress]{revtex4-2}
\usepackage{amsmath,amsfonts,amssymb}

\usepackage[inline]{enumitem}
\usepackage{float}
\usepackage{xr-hyper}
\usepackage{hyperref}
\hypersetup{colorlinks, allcolors=Blue}
\usepackage{mathtools}

\usepackage[utf8]{inputenc}

\usepackage{bibunits}
\setcitestyle{super}

\usepackage{graphicx}
\usepackage{siunitx}
\usepackage{tikz}
\usepackage{xcolor}

\newcommand{\ud}{\mathrm{d}}

\newcommand{\revision}[1]{{\color{Black}#1}}

\makeatletter
\newcommand*{\addFileDependency}[1]{
  \typeout{(#1)}
  \@addtofilelist{#1}
  \IfFileExists{#1}{}{\typeout{No file #1.}}
}
\makeatother

\newcommand*{\myexternaldocument}[2]{%
    \externaldocument{#1/#2}%
    \addFileDependency{#2.tex}%
    \addFileDependency{#1/#2.aux}%
}

\myexternaldocument{anc}{supplemental}

\begin{document}

\title{Panoramic mapping of phonon transport from ultrafast electron diffraction and machine learning}

\author{Zhantao Chen}
\thanks{Corresponding author. \\\href{mailto:zhantao@mit.edu}{zhantao@mit.edu}
\\\href{mailto:xshen@slac.stanford.edu}{xshen@slac.stanford.edu}
\\\href{mailto:mingda@mit.edu}{mingda@mit.edu}}
\affiliation{Quantum Measurement Group, MIT, Cambridge, MA 02139}
\affiliation{Department of Mechanical Engineering, MIT, Cambridge, MA 02139}
\author{Xiaozhe Shen}
\thanks{Corresponding author. \\\href{mailto:zhantao@mit.edu}{zhantao@mit.edu}
\\\href{mailto:xshen@slac.stanford.edu}{xshen@slac.stanford.edu}
\\\href{mailto:mingda@mit.edu}{mingda@mit.edu}}
\affiliation{SLAC National Accelerator Laboratory, Menlo Park, CA 94205}
\author{Nina Andrejevic}
\affiliation{Quantum Measurement Group, MIT, Cambridge, MA 02139}
\affiliation{Department of Materials Science and Engineering, MIT, Cambridge, MA 02139}
\author{Tongtong Liu}
\affiliation{Quantum Measurement Group, MIT, Cambridge, MA 02139}
\affiliation{Department of Physics, MIT, Cambridge, MA 02139}
\author{Duan Luo}
\affiliation{SLAC National Accelerator Laboratory, Menlo Park, CA 94205}
\author{Thanh Nguyen}
\affiliation{Quantum Measurement Group, MIT, Cambridge, MA 02139}
\affiliation{Department of Nuclear Science and Engineering, MIT, Cambridge, MA 02139}
\author{Nathan C. Drucker}
\affiliation{Quantum Measurement Group, MIT, Cambridge, MA 02139}
\affiliation{John A. Paulson School of Engineering and Applied Science, Harvard University, Cambridge, MA 02138}
\author{Michael E. Kozina}
\affiliation{SLAC National Accelerator Laboratory, Menlo Park, CA 94205}
\author{Qichen Song}
\affiliation{Department of Mechanical Engineering, MIT, Cambridge, MA 02139}
\author{Chengyun Hua}
\affiliation{Materials Science and Technology Division, Oak Ridge National Laboratory, Oak Ridge, Tennessee 37831}
\author{Gang Chen}
\affiliation{Department of Mechanical Engineering, MIT, Cambridge, MA 02139}
\author{Xijie Wang}
\affiliation{SLAC National Accelerator Laboratory, Menlo Park, CA 94205}
\author{Jing Kong}
\affiliation{Department of Electrical Engineering and Computer Science, MIT, Cambridge, MA 02139}
\author{Mingda Li}
\thanks{Corresponding author. \\\href{mailto:zhantao@mit.edu}{zhantao@mit.edu}
\\\href{mailto:xshen@slac.stanford.edu}{xshen@slac.stanford.edu}
\\\href{mailto:mingda@mit.edu}{mingda@mit.edu}}
\affiliation{Quantum Measurement Group, MIT, Cambridge, MA 02139}
\affiliation{Department of Nuclear Science and Engineering, MIT, Cambridge, MA 02139}

\date{\today}

\begin{abstract}
    One central challenge in understanding phonon thermal transport is a lack of experimental tools to investigate mode-based transport information. Although recent advances in computation lead to mode-based information, it is hindered by unknown defects in bulk region and at interfaces. Here we present a framework that can reveal microscopic phonon transport information in heterostructures, integrating state-of-the-art ultrafast electron diffraction (UED) with advanced scientific machine learning. Taking advantage of the dual temporal and reciprocal-space resolution in UED, we are able to reliably recover the frequency-dependent interfacial transmittance with possible extension to frequency-dependent relaxation times of the heterostructure. This enables a direct reconstruction of real-space, real-time, frequency-resolved phonon dynamics across an interface. Our work provides a new pathway to experimentally probe phonon transport mechanisms with unprecedented details.
\end{abstract}

\maketitle

The ability to efficiently transport, convert, and store thermal energy plays an indispensable role in promoting decarbonization and mitigating global warming \cite{henry2020five}. Significant efforts have been directed to understand thermal transport at the nanoscale \cite{cahill2014nanoscale} driven by applications such as thermoelectric energy harvesting \cite{Minnich2009Review}, heat management in microelectronics \cite{moore2014emerging}, high-efficiency thermal storage systems \cite{amaral2017storage}, and passive cooling of structural materials \cite{li2019radiative}. However, our understanding of phonon thermal transport is largely hindered by the lack of experimental tool that can resolve mode-based phonon transport, both in the bulk region and across interfaces. Observables like heat capacity and thermal conductivity are mode- or frequency-integrated. Although computation can resolve phonon transport modal information such as interface transmittance and mean-free-paths, it relies heavily on detailed atomic and defect configurations that are usually unknown. Here, we present an integrated experimental-computational-machine learning framework that can resolve frequency-dependent phonon interfacial transmittance and phonon relaxation times using a laser-pump, electron-probe ultrafast electron diffraction (UED) setup. The acquired information offers unprecedentedly detailed knowledge on phonon transport, enabling new understanding that will have both fundamental and practical importance.

In the past two decades, remarkable progress has been made in understanding phonon thermal transport, enabled by advances in experimental and computational techniques \cite{cahill2004analysis,broido2007intrinsic,Schmidt2008tdtr,lindsay2013phononisotope,regner2013broadband, Hu2015mfp, hua2017experimental,Maznev2011nondiff, ravichandran2018spectrally, forghani2019phonon}. Novel phonon transport regimes have been observed\cite{chen2021non-fourier,qian2021phonon}, such as quantized, ballistic phonon transport \cite{schwab2000measurement}, coherent phonon transport \cite{Luckyanova2012coherent}, phonon Anderson localization \cite{luckyanova2018phonon}, and hydrodynamic phonon transport \cite{Huberman2019science,Machida2020Science}. On the other hand, a few works have demonstrated the power of using electron and x-ray diffraction to study phonon spectroscopy\cite{trigo2010imaging, zhu2015phonon, yan2021single-defect}, suggesting that diffraction techniques may also serve as powerful tools to study frequency-dependent phonon transport. However, these techniques are limited to bulk crystalline materials. Frequency-resolved interfacial transport remains a major challenge.

\begin{figure*}[!t]
    \centering
    \includegraphics[width=0.85\linewidth]{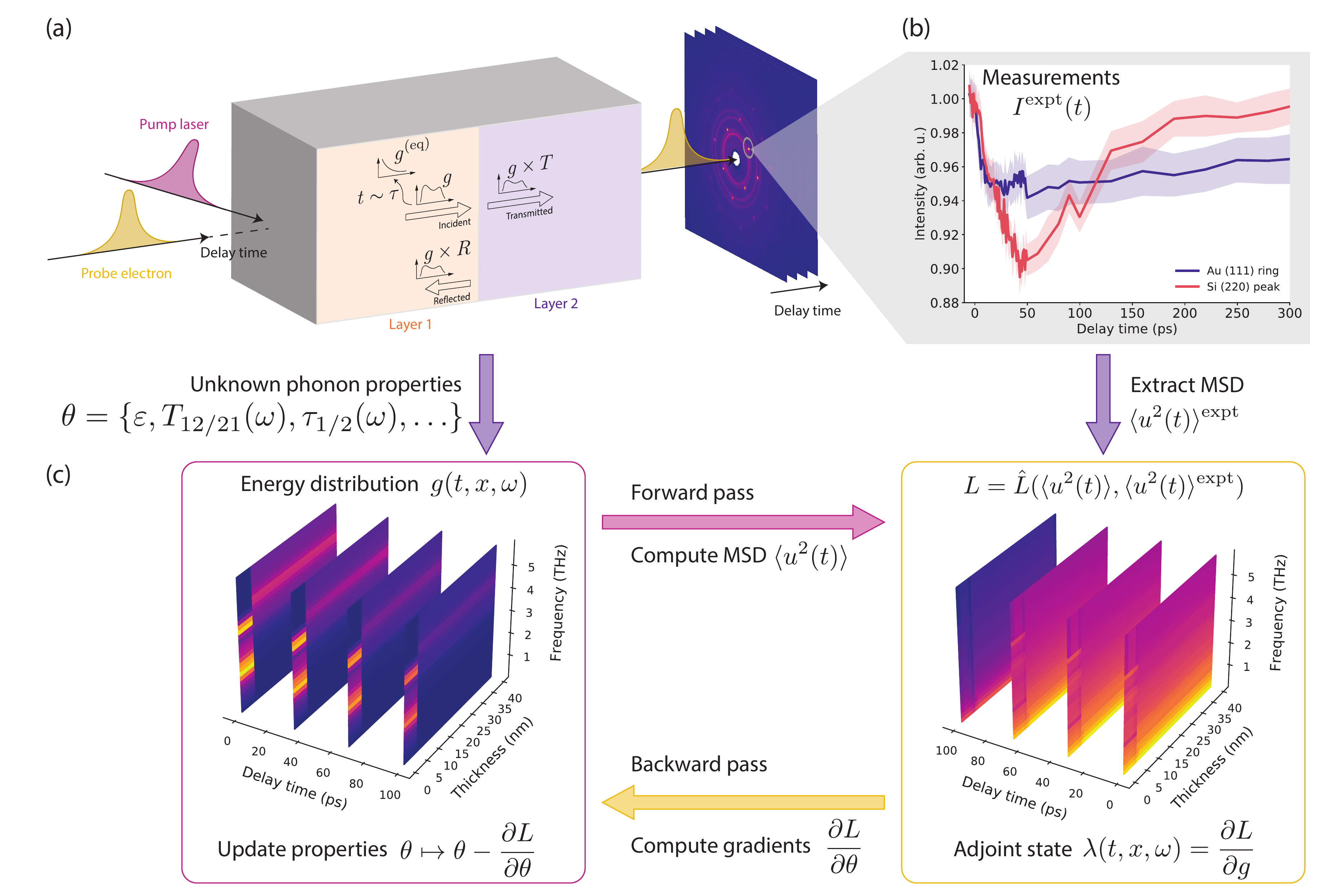}
    \caption{\textbf{Workflow for learning phonon transport properties from ultrafast electron diffraction.} \textbf{(a)} Illustration of a UED experiment on a two-layer heterostructure. A fs pump laser generates a thermal excitation, while ultrafast electrons generate the time-dependent diffraction patterns at a sequence of delay times $t$. \textbf{(b)} Representative measurements of $t$-dependent relative diffraction intensities of the two layers, which can be used to extract mean-squared displacements (MSD) $\langle u^{2}(t) \rangle^{\mathrm{expt}}$ through the atomic thermal motions from the Debye-Waller factor. \textbf{(c)} Illustration of the iterative algorithm used to learn phonon transport properties $\theta$. The phonon energy distribution $g(t,x,\omega)$ is calculated using the Boltzmann transport equation (BTE) with an initial guess for $\theta$ (which is a set of unknown phonon properties, such as phenomenological energy loss coefficients $\varepsilon_{1/2}$, frequency-dependent interfacial transmittance $T_{12/21}(\omega)$, and frequency-dependent relaxation times $\tau_{1/2}(\omega)$), and is then used to computationally obtain the MSDs $\langle u^{2}(t) \rangle$. The discrepancies between $\langle u^{2}(t) \rangle$ and the target, $\langle u^{2}(t) \rangle^{\mathrm{target}}=\langle u^{2}(t) \rangle^{\mathrm{expt}}$ here, are measured by a loss function $L$. The parameter gradients $\partial L/\partial \theta$ are obtained by the adjoint-state method and fed into an optimizer to minimize $L$ and provide improved estimates of $\theta$.}
    \label{fig:figure_1}
\end{figure*}

The UED measures the time-dependent diffraction patterns of each layer in a heterostructure with sub-ps resolution (Fig.\@ \ref{fig:figure_1}(a)). Although the diffraction intensities in principle are linked to the atomic displacements through microscopic phonon transport and local temperature variations, the inverse problem of extracting phonon transport information, such as interfacial transmittance $T_{12/21}(\omega)$, and relaxation times $\tau_{1/2}(\omega)$ from time resolved diffraction information is difficult. \revision{The difficulties mainly originate from the high-dimensional phonon dynamics that depend on time, space, and frequency, and low-dimensional observables that are only time-dependent, and layer-specific diffraction patterns contain integrated information from all atoms in the domain. Given the advances in \textit{ab initio} phonon computations\cite{minnich2015advances}, we consider some basic phonon properties such as density-of-states (DOS) and group velocities in each layer are known, and assume that the diffraction intensity smear comes solely from the lattice contributions after full electron thermalization. A number of key thermal transport parameters appeared in Boltzmann transport equation (BTE) can be reconstructed using advanced ajoint-state \cite{cao2003adjoint} and automatic differentiation machine learning techniques \cite{baydin2018automatic}, including frequency-dependent phonon transmittance across interface, with possible extension to frequency-dependent relaxation times, thereby is termed ``panoramic mapping''.} The machine learning techniques play a crucial role to reliably output high-dimensional frequency-dependent thermal transport information from diffraction spots' time evolution. Since interfaces often play a key role in thermal transport, the unprecedentedly detailed knowledge of interfacial transport could lead to various applications including enhancement of heat transfer through interfacial engineering, development of high thermal conductivity materials for improved heat transfer, and materials and architectures design for thermal energy storage. 

\section*{The framework setup}

The overall architecture of the framework leading to panoramic phonon transport mapping is summarized in Fig.\@ \ref{fig:figure_1}. First, we employ MeV-UED to acquire the time-resolved electron diffraction patterns of a heterostructure after the fs-laser excitation (pump laser) with transmission geometry (Fig.\@ \ref{fig:figure_1}(a))\cite{weathersby2015mega, shen2018femtosecond}. Diffraction spots coming from different layers in a heterostructure system offer a natural layer-resolved information. Consider an Au/Si heterostructure used in experiments, the intensity evolution of the (111) ring in poly-crystalline Au and the (220) diffraction spot in single-crystalline Si are 
shown in Fig.\@ \ref{fig:figure_1}(b) as an illustration. Our goal is to use such time-dependent diffraction information to extract the phonon transport properties. The key challenge is that diffraction results from contribution of all atoms in the layer while phonon transport depends on both time and space. Using the diffraction information of both layers in reciprocal space increases the reliability for information extraction. The Debye-Waller factor links the diffraction intensity $I(t)$ to the mean-squared atomic displacements (MSD) $\langle u^{2}(t)\rangle$ as $I(t)\propto \exp(-\frac{1}{3}q^{2}\langle u^{2}(t)\rangle)$ ($q=|\mathbf{q}|$ is the norm of the electron-beam wavevector transfer, $\langle\ldots\rangle$ indicates an ensemble average over atoms) at each time $t$. By feeding a set of generic phonon transport properties $\theta$ (e.g.\@ transmittance), $\langle u^{2}(t)\rangle$ is directly computable from BTE, i.e., each $\langle u^{2}(t)\rangle$ time-series are labeled by parameters $\theta$. To extract the phonon transport from a target data of thickness-averaged MSD $\langle u^{2}(t)\rangle^{\mathrm{target}}$, which can come from real experimental data, we solve an optimization problem by minimizing the mean absolute loss $L=\frac{1}{N}\sum_{n=1}^{N}|\langle u^{2}(t_{n})\rangle-\langle u^{2}(t_{n})\rangle^{\mathrm{target}}|$ between the computed and experimentally extracted MSDs.

\begin{figure*}[!ht]
    \centering
    \includegraphics[width=0.85\linewidth]{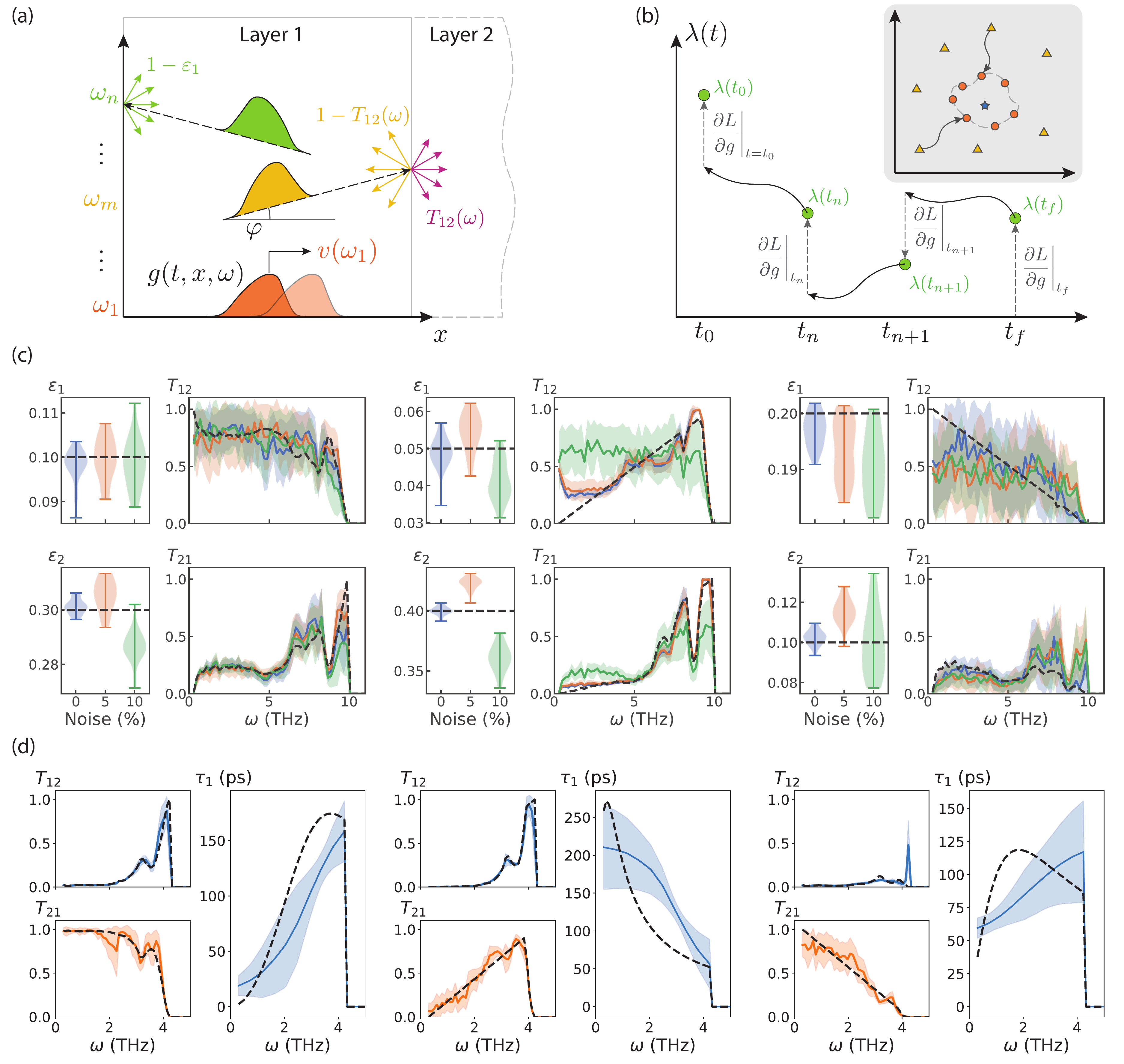}
    \caption{\textbf{The forward model, optimization process, and performance of the proposed framework on synthetic data.} \textbf{(a)} Schematic illustration of the theoretical model. Each phonon mode of frequency $\omega$ propagates its energy (characterized by the distribution $g(t,x,\omega)$) at its group velocity $v(\omega)$ obeying the BTE. At the material boundary of layer 1, only partial energy ($1-\varepsilon_{1}$) gets diffusely back scattered, while the rest dissipates into the environment. At the interface between two layers, the energy from layer 1 is transmitted (reflected) diffusely with coefficients $T_{12}(\omega)$ ($1-T_{12}(\omega)$). \textbf{(b)} Illustration of the adjoint-state method, where the continuous backward trajectories (solid curves with leftward arrows) indicate evolutions induced by $\lambda\frac{\partial h}{\partial \lambda}$ in the Eq.\@ \eqref{eqn:adj_eqns}, the dashed vertical lines correspond to sudden changes due to the presences of discrete measurement data points and correspond to $\frac{\partial L}{\partial g}$ (the illustration is adapted from Fig.\@ 2 in the Ref.\@ \citenum{chen_neural_2018}). Inset: simultaneous optimization of independently-initialized phonon parameters. \textbf{(c)} Representative reconstructions of $\varepsilon_{1/2}$ and transmittance $T_{12/21}(\omega)$ in the absence of noise (blue) the presence of $\delta=0.05,0.1$ noise (orange and green, respectively) of a $5\,\mathrm{nm}$ Al\textsubscript{2}O\textsubscript{3} and $35\,\mathrm{nm}$ Al heterostructure. The Al\textsubscript{2}O\textsubscript{3} and Al layers correspond to the subscripts 1 and 2, respectively. \textbf{(d)} Representative reconstructions for transmittance and relaxation times from synthetic MSDs of a $5\,\mathrm{nm}$ Au and $35\,\mathrm{nm}$ Si heterostructure. The Au and Si layers correspond to subscripts 1 and 2, respectively. The solid lines and shaded areas represent the mean reconstructions and those within one standard deviation, respectively. The dashed lines indicate true values. The statistics presented here is obtained from 20 sets of independently initialized initial parameters.}
    \label{fig:figure_2}
\end{figure*}

The entire process of minimizing loss function $L$ and extracting phonon properties $\theta$ is composed of two parts, the forward pass and the backward pass (Fig.\@ \ref{fig:figure_1}(c)). In the forward pass, we simulate the MSD $\langle u^{2}(t)\rangle$ given an initial guess of the parameters $\theta$; the theoretical framework is visualized in Fig.\@ \ref{fig:figure_2}(a). The time-dependent local MSD $\langle u^{2}(t,x)\rangle$ is directly linked to the phonon energy distribution  $g(t,x,\omega) $ as \cite{als2011elements}
\begin{equation}\label{eqn:msd_tx}
    \langle u^{2}(t,x)\rangle = \frac{1}{N m}\int_{0}^{\infty}\frac{1}{\omega^{2}}\left(g(t,x,\omega)+\frac{\hbar \omega}{2}D(\omega)\right)\  \ud \omega,
\end{equation}
where $D(\omega)$, $N$, and $m$ are the phonon DOS, number of atoms per unit cell, and the atomic mass for monoatomic solid, respectively (a more general formulation is provided in Supplementary Information \ref{SI_sec:DW_MSDs}). The measured displacements $\langle u^{2}(t)\rangle$ from UED is the thickness-averaged displacement, that $\langle u^{2}(t)\rangle=\frac{1}{L}\int_{0}^{L}\langle u^{2}(t,x)\rangle\ \ud x$ due to the transmission geometry. Due to the much larger spot size of the pump laser ($300\,\mathrm{\mu m}$ full width at half maximum, FWHM) compared to that of the probe electron ($100\,\mathrm{\mu m}$ FWHM), the changing diffraction intensities can be attributed to the cross-plane phonon transport process, which can be well captured by a one-dimensional BTE under the relaxation time approximation (RTA), that
\begin{equation}
\begin{split}
    \frac{\partial g(t,x,\omega)}{\partial t} &+ \mu v(\omega)\frac{\partial g(t,x,\omega)}{\partial x} = \\ 
    &-\frac{g(t,x,\omega)-g^{\mathrm{eq}}(T(t,x),\omega)}{\tau(\omega)},
\end{split}
\end{equation}
where $v(\omega)$ and $\tau(\omega)$ denote the frequency-dependent group velocities and relaxation times, respectively, and $\mu=\cos(\varphi)$ projects the group velocity at an angle $\varphi\in[0,\pi]$ to the $x$-direction (Fig.\@ \ref{fig:figure_2}(a)). 

\revision{The phonon transport across the interface is dictated by transmittance. For example, only partial energy of the right-propagating phonon in layer 1, $0\le T_{12}(\omega)\le 1$, gets transmitted to layer 2 while the rest get reflected, as illustrated in Fig.\@ \ref{fig:figure_2}(a). The energy exchange rate between the phonon system and the environment is phenomenologically parameterized by $\varepsilon$ ($0\le \varepsilon\le 1$) in the boundary conditions (Fig.\@ \ref{fig:figure_2}(a)). More precise formulations of the above processes are provided in the Methods section as well as in Supplementary Information \ref{SI_sec:boltzmann_transport_equations}. In addition, we also considered another formulation that takes care of energy loss through bulk of materials in Supplementary Information Section \ref{SI_sec:bulk_energy_loss}.}

In the backward pass, we make use of the adjoint equations of BTE to obtain the loss function gradients necessary in any gradient-based optimization to refine the initial guess of the transport parameters, $\theta$. In particular, the loss function gradient $\partial L/\partial\theta$ is solved by integrating the following equations backward in time from the final simulation moment $t=t_{f}$ to the initial one $t_{0}$
\begin{equation}\label{eqn:adj_eqns}
\begin{aligned}
    \frac{\partial \lambda}{\partial t} &= - \lambda\frac{\partial h}{\partial g} + \frac{\partial L}{\partial g},\qquad &\lambda(t_{f})=0,\\
    \frac{\partial \eta}{\partial t} &= -\lambda\frac{\partial h}{\partial \theta},\qquad &\eta(t_{f})=0,
\end{aligned}
\end{equation}
where $h=\partial g / \partial t$, and $\lambda = \partial L/\partial g$ is the adjoint state of the energy distribution function $g$ \revision{whose backward time-evolution collects discrepancies between simulated MSDs and experimental MSDs (Fig.\@ \ref{fig:figure_2}(b)}). The $\eta(t_{0})$ represents the overall gradient counting all measurements within time range $[t_{0},t_{f}]$ (Supplementary Information \ref{SI_sec:optimization_method}). The combination of adjoint-state method and automatic differentiation has demonstrated huge success in solving large number of neural differential equations and extracting information, even for ill-posed problems with multiple local minima\cite{chen_neural_2018}. In the current case, these machine learning techniques enable a reliable parameter extraction from BTE solutions and UED measurements. More details are presented in Methods section and Supplementary Information \ref{SI_sec:optimization_method}.

\revision{In practice, we independently initialize a set of phonon properties, $\{\theta_{1}^{(0)},\theta_{2}^{(0)}\ldots,\theta_{N}^{(0)}\}$, and update them simultaneously to some final reconstructed phonon properties $\{\theta_{1}^{(\mathrm{f})},\theta_{2}^{(\mathrm{f})}\ldots,\theta_{N}^{(\mathrm{f})}\}$. The population of reconstructions is able to better overcome local minima of the loss function with their averages $\bar{\theta}^{(\mathrm{f})}=\frac{1}{N}\sum_{n=1}^{N}\theta_{n}^{(\mathrm{f})}$, and can quantify reconstruction uncertainties with their standard deviations, as illustrated in the inset panel of Fig.\@ \ref{fig:figure_2}(b) and further discussed in the Method section. Since $T_{12}(\omega_{n})$ and $T_{21}(\omega_{n})$ are connected by the detailed balance (Method section), we choose to optimize over an composite function $\mathcal{T}(\omega_{n})=\max(T_{12}(\omega_{n}), T_{21}(\omega_{n}))$ for each sampled frequency point $\omega_{n}$.}

\section*{Numerical Verifications}

\revision{The ground-truth transmittance of a real experimental sample are largely inaccessible by other experimental methods.} We first benchmark our framework over numerically simulated MSDs (synthetic data) with known ground truth $\theta^{\mathrm{gt}}$ on a $5\,\mathrm{nm}$ Al\textsubscript{2}O\textsubscript{3} and $35\,\mathrm{nm}$ Al heterostructure. As shown in Fig.\@ \ref{fig:figure_2}(c), comparisons between $\bar{\theta}^{(\mathrm{f})}$ (solid lines for averages and shaded areas for standard deviations) and $\theta^{\mathrm{gt}}$ (dashed lines) are made for three representative sets of phonon transmittance $T_{12/21}(\omega)$ (right panels) and material-air boundary energy loss coefficients $\varepsilon_{1/2}$ (left panels). Five sets of MSD with distinct initial conditions are used to perform the reconstructions (Supplementary Information Table \ref{SI_tab:initial_conditions_Al2O3_Al}). Excellent agreements are obtained for both reconstructions of the $T_{12/21}(\omega)$ and $\varepsilon_{1/2}$. We notice that the extracted transmittance are spread around $\mathcal{T}^{\mathrm{gt}}$ with larger standard deviations. This behavior can be understood by noting that transmittance only affects a portion of phonon energies in different layers but has no direct influence on the total energy in the heterostructure, while the $\varepsilon$'s directly determine the energy exchange rate with the external environment at material boundaries and dominates the overall trend of $\langle u^{2}(t)\rangle$.  The comparatively lower impact of transmittance leads to slower convergence during the learning process compared to that of energy loss coefficients. However, after sufficient number of training epochs, our framework can successfully capture these nuanced properties, as the averaged reconstructions faithfully reflect key features of the true profiles. 

To prepare for realistic experimental extraction, we simulate noisy measurements by perturbing the computed MSD according to $\langle u^{2}(t_{n})\rangle_{\mathrm{noisy}}^{\mathrm{target}}=\delta(t_{n}) \langle u^{2}(t_{n})\rangle^{\mathrm{target}}$, where the prefactor $\delta(t_{n})$ at each time point $t_{n}$ is randomly drawn from a uniform distribution  $\delta(t_{n})\sim\mathcal{U}[1-\delta,1+\delta]$ with $0\le\delta\le1$ representing the noise level. In Fig.\@ \ref{fig:figure_2}(c), we demonstrate three representative sets of  reconstructions, where different colors (blue, orange, green) represent the reconstructions under different levels of noises ($\delta=0\%,\, 5\%,\, 10\%$). We see that even with broader spread of the retrieved properties with increased noise level $\delta$ grows, reasonable agreements between reconstructed and ground truth values are still largely maintained. This suggests that our framework is capable of accommodating moderate noises.

\begin{figure*}[!ht]
    \centering
    \includegraphics[width=0.85\linewidth]{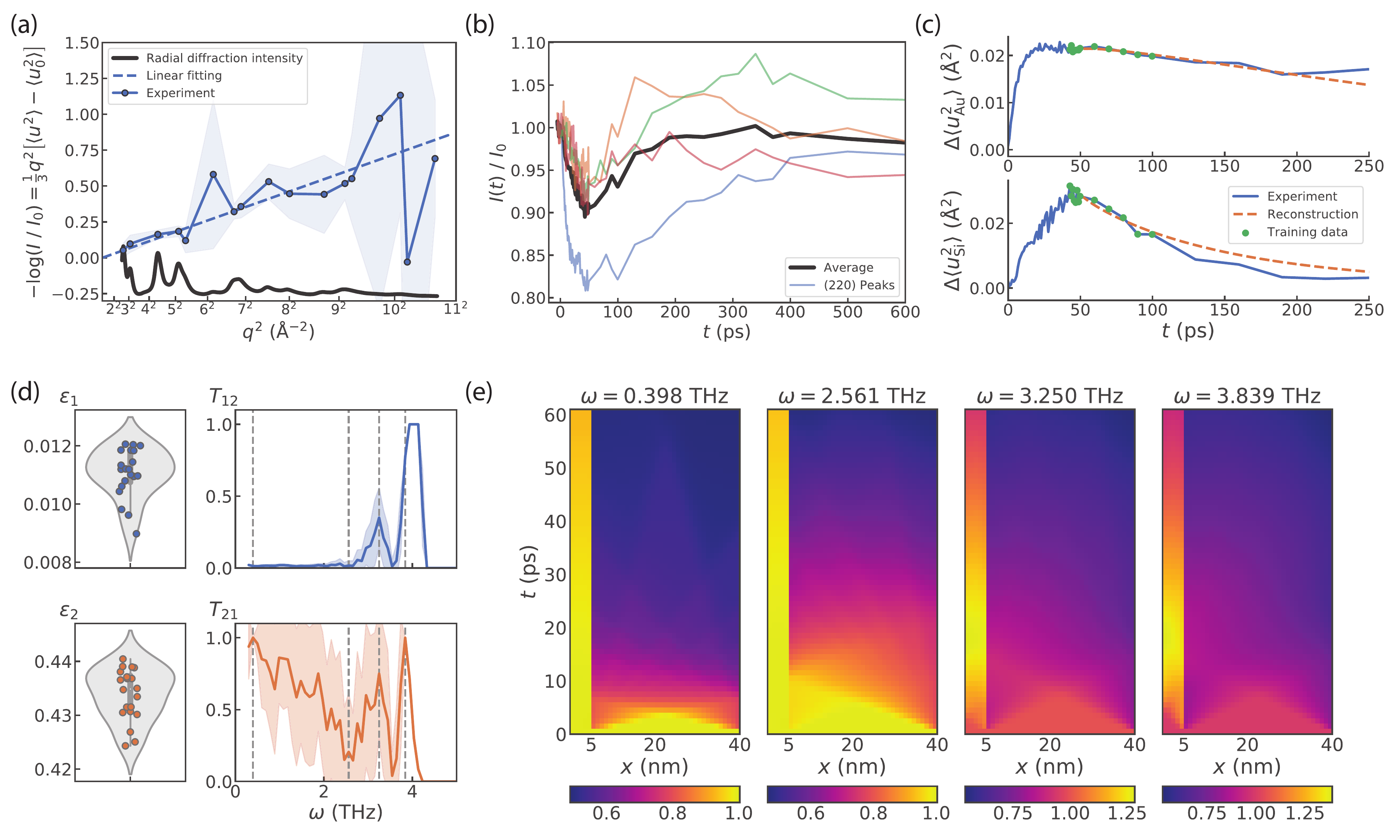}
    \caption{\textbf{Reconstructions of phonon transport properties from experimentally-measured UED of the Au/Si heterostructure.} \textbf{(a)} Representative linear fitting of $-\log(I/I_{0})$ versus $q^{2}$ to extract the MSD from the Au diffraction ring intensities at a particular delay time moment. The $\langle u_{0}^{2}\rangle$ and $I_{0}$ represent room-temperature MSD and diffraction intensity, respectively. This process can be repeated for every collected time point to obtain the complete time-dependent $\langle u_{\mathrm{Au}}^{2}(t)\rangle$. \textbf{(b)} Time-dependent diffraction intensities of Si at $(220)$ and equivalent $\mathbf{q}$-points (transparent thin lines). The MSD of Si is extracted from the averaged intensities (the thick black line). \textbf{(c)} Comparisons between measured and reconstructed MSDs of the Au (top panel) and Si (bottom panel) layers as functions of the delay time. \textbf{(d)} Reconstructed energy loss coefficients and transmittance. The statistics presented here is obtained from 20 sets of independently initialized parameters. \textbf{(e)} Evolution of normalized phonon energy distributions at selected frequencies, $\int_{-1}^{1}g(t,x,\omega,\mu)\ \ud\mu / \int_{-1}^{1}g(t_{0},x,\omega,\mu)\ \ud\mu$. The frequencies of four panels from left to right correspond to the same order of dashed vertical lines shown in part (d). The Au/Si interface is located at $x=5\,\mathrm{nm}$.}
    \label{fig:figure_4}
\end{figure*}

With the success of Al/Al$_2$O$_3$ noise test, we further study a 5nm Au/35nm Si heterostructure, which is the same material used in experiments. We can further extend the proposed framework to extract phonon relaxation times. However, we note that the reconstruction of relaxation times is typically more challenging due to short signal time span and even lower gradient magnitudes (Supplementary Information Fig.\@ \ref{fig:SI_gradients}). This indicates that better knowledge about energy loss coefficients and transmittance are desirable to extract relaxation times from experiments. For simplicity, here we assume that the energy loss coefficients of both layers and the relaxation time of the second layer in a heterostructure are known, and perform simultaneous recovery on transmittance $\mathcal{T}(\omega)$ on both layers and relaxation times $\tau(\omega)$ of the first layer. In Fig.\@ \ref{fig:figure_2}(d), we show three representative reconstructions of transmittance $\mathcal{T}(\omega)$ and relaxation times $\tau_{\mathrm{Au}}(\omega)$ from synthetic MSDs of the $5\,\mathrm{nm}$ Au/$35\,\mathrm{nm}$ Si heterostructure. More details about reconstructions of $\mathcal{T}(\omega)$ and $\tau(\omega)$ are discussed in Supplementary Information \ref{SI_subsec:recon_T_tau}.

\section*{Application to experimental data}

We conduct UED measurements on a heterostructure composed of $5\,\mathrm{nm}$ thick polycrystalline Au and $35\,\mathrm{nm}$ thick (001)-oriented single crystal Si membrane fabricated with micro-electromechanical system (MEMS) at MeV-UED beamline at LCLS, SLAC National Accelerator Laboratory. We obtain the time-resolved diffraction patterns consisting of diffraction rings (Au) and Bragg spots (Si) (see a representative example in Supplementary Information Fig.\@ \ref{fig:SI_schematic_spectrum_diffraction}). For each diffraction pattern at a given time delay $t$, the MSDs of each layer can be extracted separately. For the Au layer, we use the full $q$-dependent intensities to perform linear fitting between $-\log(I(t)/I_{0})$ and $q^{2}$ (Fig.\@ \ref{fig:figure_4}(a) shows the linear fitting for a particular delay time moment), which can be repeated over all measured time points to obtain $\langle u_{\mathrm{Au}}^{2}(t) \rangle$. While for the Si layer, diffraction spot intensities from the $(220)$ family with high signal-to-noise ratio are used (Fig.\@ \ref{fig:figure_4}(b)). We attribute the asymmetric diffraction intensities (shown by the four transparent lines in Fig.\@ \ref{fig:figure_4}(b)) observed in the equivalent $\mathbf{q}$-points of Si to a possibly rugged Au/Si interface due to mismatches of in-plane thermal strains at two sides after the pump laser excitation\cite{harb2009excitation}, which shall not affect the cross-plane transport. The resulting MSDs are plotted as solid lines in Fig.\@ \ref{fig:figure_4}(c). More details about experiment data analysis are presented in Supplementary Information \ref{SI_sec:experiment_method}. 

The proposed framework can then be applied to learn transmittance from experimental MSDs. In particular, given the existence of plateau in the $\langle u_{\mathrm{Au}}^{2}(t)\rangle$, the initial time $t_{0}$ is chosen to be the moment when $\langle u_{\mathrm{Si}}^{2}(t)\rangle$ reaches maximum, which occurs at $t_{0}\approx 42.8\,\mathrm{ps}$. The initial phonon energy distribution is taken to be at equilibrium, namely $g(t_{0},x,\omega,\mu)=g^{\mathrm{eq}}(t_{0},x,\omega)\equiv \frac{1}{4\pi}\hbar\omega D(\omega)f_{\mathrm{BE}}(T(t_{0},x))$ for each layer, with the initial temperature $T(t_{0},x)$ being uniform in space $x$. The initial temperatures of each layer are solved using Eq.\@ \eqref{eqn:msd_tx} to match $\langle u^{2}(t_{0})\rangle$, resulting in $T_{\mathrm{Au}}\approx 495.6\,\mathrm{K}$ and $T_{\mathrm{Si}}\approx 610.3\,\mathrm{K}$. 

The subsequent MSDs collected over the total time of $t_{f}-t_{0}\approx 57\,\mathrm{ps}$, indicated by circular markers in Fig.\@ \ref{fig:figure_4}(c), serve as the target $\langle u^{2}\rangle^{\mathrm{target}}$ to obtain the reconstructed transmittance displayed in Fig.\@ \ref{fig:figure_4}(d). Due to the lack of ``ground truth'' parameters of the experiment data, we are unable to directly quantify the reconstruction performance. However, numerical experiments can be performed with synthetic MSDs prepared at exactly the same delay time points, initial temperatures (training samples), and comparable noise level, building a level of confidence (Supplementary Information  \ref{SI_subsec:numexp_AuSi}). Due to the limited data points and a single training sample in experimental data, our framework was not able to achieve full reconstruction performance as done in noisy synthetic data shown in Fig.\@ \ref{fig:figure_2}. Even so, the reconstruction on transmittance can still shed light on the interfacial thermal transport with a fine frequency-dependent knowledge.

With the phonon transmittance reconstructued in Fig.\@ \ref{fig:figure_4}(d), we are then able to reveal the real-time, real-space, and frequency-dependent dynamics of the phonon energy flow in the heterostructure (Fig.\@ \ref{fig:figure_4}(e)). Due to the relatively high surface energy loss coefficients of the Si layer ($\bar{\varepsilon}_{\mathrm{Si}}\approx 0.43$), phonon energies decrease rapidly at the boundary $x=40\,\mathrm{nm}$ through boundary scattering. The relative magnitudes of the mode-integrated energy distributions at the interface ($x=5\,\mathrm{nm}$) of both layers indicate a positive energy flux from the Si layer to the Au layer, as dictated by $T_{21}(\omega)$, resulting in a slightly increased lattice temperature of the Au shortly after $t_{0}$. As to the Au layer, the energy distribution $g_{\mathrm{Au}}(t,x,\omega)$ is spread more uniformly, due to its smaller thickness, lower energy loss coefficients ($\bar{\varepsilon}_{\mathrm{Au}}\approx 0.01$) and effect from transmittance $T_{12}(\omega)$. Valuable insight about microscopic phonon transport is gained from the frequency-resolved evolution of the solid angle-integrated, normalized energy distribution, $\int_{-1}^{1}g(t,x,\omega,\mu)\ \ud\mu / \int_{-1}^{1}g(t_{0},x,\omega,\mu)\ \ud\mu$, shown in Fig.\@ \ref{fig:figure_4}(e), where we observe that the evolution of energy distributions of different phonon mode vary significantly, especially at the Au/Si interface. These and other quantitative analyses are made possible by solving the BTE using the reconstructed parameters, which can provide key insights for engineering phonon transport in diverse materials systems.

\section*{Discussion and Conclusion}

This framework offers a new avenue that can efficiently and comprehensively characterize layer-specific and frequency-resolved thermal transport properties with two key features. One is the additional reciprocal space information, where different layers result in different diffraction spots. This leads to a higher-dimensional input in $(\mathbf{q},t)$ space and enables a simultaneous reconstruction of phonon properties with less sensitivity against noise. The other feature is the capability to capture real phonon dynamics. The initial time can be chosen well after the pulse excitation and full thermalization of hot electrons without hampering information extraction, and the analysis of Debye-Waller factor focuses on the atomic displacements with minimal interference of the electronic degrees of freedom. 

There are still a few improvements and generalization can be done based on this framework. First, the validation of the extracted parameters from experimental is challenging. The partial information agreement with experimental measurements and the high-quality benchmark with computational data build a level of confidence, yet additional independent experimental validation will build further confidence on the extracted information, such as parameters like $T_{12}(\omega)$. Second, we have assumed that there is no coupling between phonons with different frequencies or inelastic scattering. This allows us to treat each phonon frequency channel independently but limits its application in the strongly anharmonic regime. Future incorporation on anharmonic effects are feasible with further modification of the framework. Third, we have assumed that the phonon DOS $D(\omega)$ and group velocities $v(\omega)$ are known, yet in nanostructures, they may be subject to change due to size effect. We note that the drastic size effect is only apparent for very thin, few-nm-thick films \cite{Balandin2005}, and even in that regime, $D(\omega)$ and $v(\omega)$ are still computable with relatively high fidelity. The effect of $D(\omega)$ and $v(\omega)$ are subject to further investigation. 

In this work, we develop a machine learning-informed computational framework to analyze the time-resolved diffraction patterns in UED to infer frequency-resolved interfacial thermal transport at the nanoscale. The same principle is also applicable to other time-resolved diffraction experiments. The combination of the adjoint-state and automatic differentiation machine learning approach makes the BTE model differentiable with respect to its phonon properties without compromising physical validity. We demonstrate its power in reliably learning multiple phonon properties in distinct scenarios, as well as its robustness against measurement noise. Our approach opens up a new way to study frequency-resolved nanoscale thermal transport with foreseeable impact in the areas of improving thermal management to realize higher energy efficiency, facilitate sustainable usage of natural resources, and reduce global warming. Given the generality of automatic differentiation and the adjoint-state method for solving differential-equation based physical models, we anticipate the framework could benefit the study of a wide range of physical systems, enabling the acquisition of hidden information from complex observables in a dynamical system and informing the design of better measurement strategies.

\section*{Methods}
\noindent\textbf{Forward solving and backward parameter extraction for BTE}\\

To establish the forward model to solve the BTE, we need boundary conditions, initial conditions and detailed balance conditions. The mismatch of vibrational properties at the interface allows only partial transmission of the phonon energy across the interface while the rest is reflected. Therefore, the energy flux $\mathcal{F}^{\pm}(t,x,\omega)=\int_{\mu\gtrless 0}\mu v(\omega)g(t,x,\omega,\mu)\ \ud\mu$ at the interface $x=x_{\mathrm{intf}}$ can be described by the interfacial boundary condition
\begin{equation}
\begin{split}
    \mathcal{F}_{2}^{+}(t,\omega)&=T_{12}(\omega)\mathcal{F}_{1}^{+}(t,\omega)+[1-T_{21}(\omega)]\mathcal{F}_{2}^{-}(t,\omega),\\
    \mathcal{F}_{1}^{-}(t,\omega)&=[1-T_{12}(\omega)]\mathcal{F}_{1}^{+}(t,\omega)+T_{21}(\omega)\mathcal{F}_{2}^{-}(t,\omega),
\end{split}
\end{equation}
where the subscripts $1,\,2$ denote the two sub-layers in the heterostructure; for instance $T_{12}(\omega)$ is the transmittance from layer 1 to 2. Since we only analyze $q$-dependent diffraction signals, the wavevector dependencies are not taken into account and mode conversion among different phonon polarizations is neglected for simplicity. Meanwhile, we restrict our attention to elastic phonon scatterings, the two sets of transmittance $T_{12}(\omega)$ and $T_{21}(\omega)$ are connected by the detailed balance \cite{swartz1989thermal}, $T_{12}(\omega)v_{1}(\omega)D_{1}(\omega)=T_{21}(\omega)v_{2}(\omega)D_{2}(\omega)$. Meanwhile, phonons are assumed to be diffusely backscattered at material-air boundaries with energy loss parameterized by $\varepsilon_{1}$ and $\varepsilon_{2}$, respectively\cite{hua2015semi}. We note that the $\varepsilon_{1}$ and $\varepsilon_{2}$ here only serve as phenomenological descriptions of the energy exchanging rate between the phonon system and the external environment. In addition to interface and boundary conditions, We choose initial condition $g(t_{0},x,\omega)=\frac{1}{4\pi}\hbar\omega D(\omega)(f_{\mathrm{BE}}(T)-f_{\mathrm{BE}}(T_{0}))$ at $t=t_{0}$, a moment after full thermalization of hot electrons. This is feasible since in an UED setup we can simply choose a $t_0$ after the maximum $\langle u^{2}\rangle$ is reached. In addition, the initial temperature profile is assumed to be spatially uniform with $T(t_{0},x)\equiv T(t_{0})$. A more detailed formulation is presented in Supplementary Information \ref{SI_sec:boltzmann_transport_equations}.

In the backward loop, one advantage of using the adjoin-state method is that there is large freedom to choose which parameters as known and which parameters are unknown and to be extracted. Since the phonon DOS $D(\omega)$ and group velocities $v(\omega)$ can be reliably computed from \textit{ab initio} methods\cite{minnich2015advances}, they are considered as known. Our discussions thus focus on reconstructing other phonon transport properties that are more challenging to obtain by conventional methods, such as frequency-dependent transmittance. Though phonon branch dependencies can be taken into account by our formalism, we restrict our attention to branches-averaged phonon transport in this work. In particular, we make use of the weighted average group velocities and relaxation times with DOS of each branch, for example, $v(\omega)=\sum_{r}D_{r}(\omega)v_{r}(\omega)/\sum_{r}D_{r}(\omega)$. The detailed-balance constraint between $T_{12}(\omega)$ and $T_{21}(\omega)$ allows us to choose the training parameters $\mathcal{T}(\omega_{n})=\max(T_{12}(\omega_{n}), T_{21}(\omega_{n}))$ at each frequency $\omega_{n}$. The $\varepsilon\in[0,1]$ could vary between samples and is also considered as unknown. Provided the minimum amount of prior knowledge, the fitting parameters $\theta$ is randomly initialized (Supplementary Information \ref{SI_subsec:init_params}). Given the high dimensionality of fitting parameters $\theta$, the loss landscape could be extremely complex and non-convex, which typically results in $\theta^{(0)}$-dependent final predictions $\theta^{(\mathrm{f})}$. To address this issue, we simultaneously optimize a set of independently-initialized parameters $\{\theta_{1}^{(0)},\theta_{2}^{(0)}\ldots,\theta_{N}^{(0)}\}$, and the final predictions are obtained by averaging over the ensemble, for example, $\bar{\theta}^{(\mathrm{f})}=\frac{1}{N}\sum_{n=1}^{N}\theta_{n}^{(\mathrm{f})}$. The benefit of this approach can be intuitively understood as approaching the ground truth $\theta^{\mathrm{gt}}$ with a population $\{\theta_{n}^{(\mathrm{f})}\}$ from different directions, schematically illustrated in Fig.\@ \ref{fig:figure_2}(b). While each cluster may be trapped in a local minimum around $\theta^{\mathrm{gt}}$, the averaged $\bar{\theta}^{(\mathrm{f})}$ can generally cancel the influence of individual local minima and thus provide a better estimate.

\begin{acknowledgments}
Z.C., N.A. and M.L. thank K. Persson for helpful discussions. Z.C. and N.A. are partially supported by U.S. Department of Energy (DOE), Office of Science, Basic Energy Sciences (BES), award No. DE-SC0021940. N.A. acknowledges the support of the National Science Foundation (NSF) Graduate Research Fellowship Program under Grant No.\@ 1122374. T.N. acknowledges the support from Sow-Hsin Chen Fellowship. T.L. and T.N. acknowledge the support from Mathworks Fellowship. M.L. is partially supported by NSF DMR-2118448 and Norman C. Rasmussen Career Development Chair, and acknowledges the support from Dr. R. Wachnik. The experiment was performed at SLAC MeV-UED and supported in part by the U.S. Department of Energy (DOE) Office of Science, Office of Basic Energy Sciences, SUF Division Accelerator \& Detector R\&D program, the LCLS Facility, and SLAC under contract Nos. DE-AC02-05CH11231 and DE-AC02-76SF00515.
\end{acknowledgments}

\bibliography{references}
\end{document}


\title{Panoramic mapping of phonon transport from ultrafast electron diffraction and machine learning\\
Supplementary Information}

\author{Zhantao Chen}
\affiliation{Quantum Measurement Group, MIT, Cambridge, MA 02139}
\affiliation{Department of Mechanical Engineering, MIT, Cambridge, MA 02139}
\author{Xiaozhe Shen}
\affiliation{SLAC National Accelerator Laboratory, Menlo Park, CA 94205}
\author{Nina Andrejevic}
\affiliation{Quantum Measurement Group, MIT, Cambridge, MA 02139}
\affiliation{Department of Materials Science and Engineering, MIT, Cambridge, MA 02139}
\author{Tongtong Liu}
\affiliation{Quantum Measurement Group, MIT, Cambridge, MA 02139}
\affiliation{Department of Physics, MIT, Cambridge, MA 02139}
\author{Duan Luo}
\affiliation{SLAC National Accelerator Laboratory, Menlo Park, CA 94205}
\author{Thanh Nguyen}
\affiliation{Quantum Measurement Group, MIT, Cambridge, MA 02139}
\affiliation{Department of Nuclear Science and Engineering, MIT, Cambridge, MA 02139}
\author{Nathan C. Drucker}
\affiliation{Quantum Measurement Group, MIT, Cambridge, MA 02139}
\affiliation{John A. Paulson School of Engineering and Applied Science, Harvard University, Cambridge, MA 02138}
\author{Michael E. Kozina}
\affiliation{SLAC National Accelerator Laboratory, Menlo Park, CA 94205}
\author{Qichen Song}
\affiliation{Department of Mechanical Engineering, MIT, Cambridge, MA 02139}
\author{Chengyun Hua}
\affiliation{Materials Science and Technology Division, Oak Ridge National Laboratory, Oak Ridge, Tennessee 37831}
\author{Gang Chen}
\affiliation{Department of Mechanical Engineering, MIT, Cambridge, MA 02139}
\author{Xijie Wang}
\affiliation{SLAC National Accelerator Laboratory, Menlo Park, CA 94205}
\author{Jing Kong}
\affiliation{Department of Electrical Engineering and Computer Science, MIT, Cambridge, MA 02139}
\author{Mingda Li}
\affiliation{Quantum Measurement Group, MIT, Cambridge, MA 02139}
\affiliation{Department of Nuclear Science and Engineering, MIT, Cambridge, MA 02139}

\date{\today}

\maketitle

\setcounter{figure}{0}
\setcounter{equation}{0}
\setcounter{table}{0}
\setcounter{section}{0}
\renewcommand{\thetable}{S\arabic{table}}
\renewcommand{\thefigure}{S\arabic{figure}}
\renewcommand{\theequation}{S\arabic{equation}}
\renewcommand{\thesection}{S\arabic{section}}

\input{anc/SI_methods}
\clearpage
\input{anc/SI_boltzmann_transport_equations}
\input{anc/SI_finite_volume_method}
\input{anc/SI_debye_waller}
\input{anc/SI_optimization_method}
\clearpage
\input{anc/SI_more_recon_tasks}
\clearpage
\input{anc/SI_AuSi_numexp}

\clearpage
\bibliography{references}

%% file: anc/SI_methods.tex
\section{Experiment Method}
\label{SI_sec:experiment_method}

\subsection{Megaelectron-volt ultrafast electron diffraction}

\begin{figure}
     \centering
     \subfloat[]{\includegraphics[width = 0.6\textwidth]{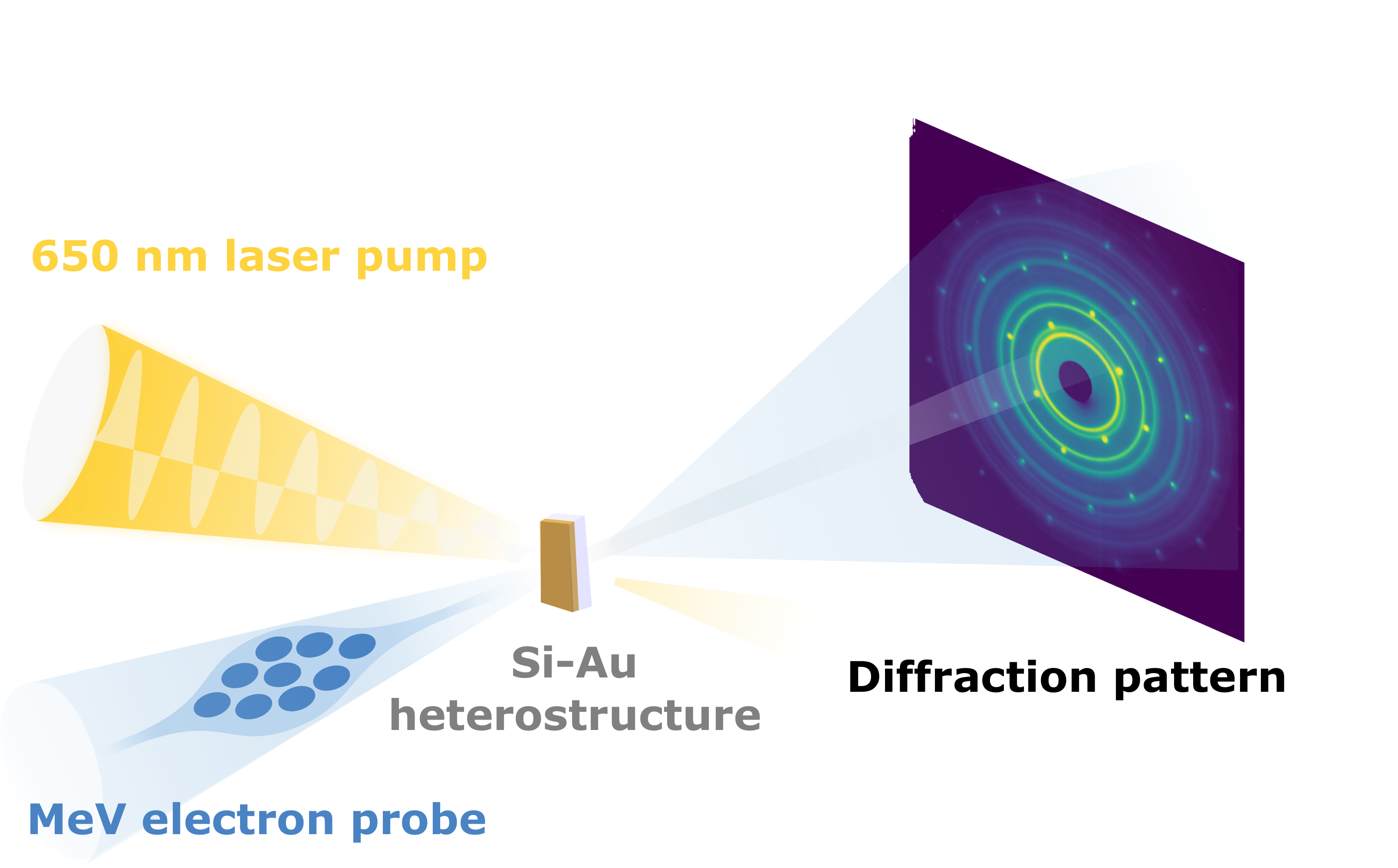}}\\
     \subfloat[]{\includegraphics[width = 0.46\textwidth]{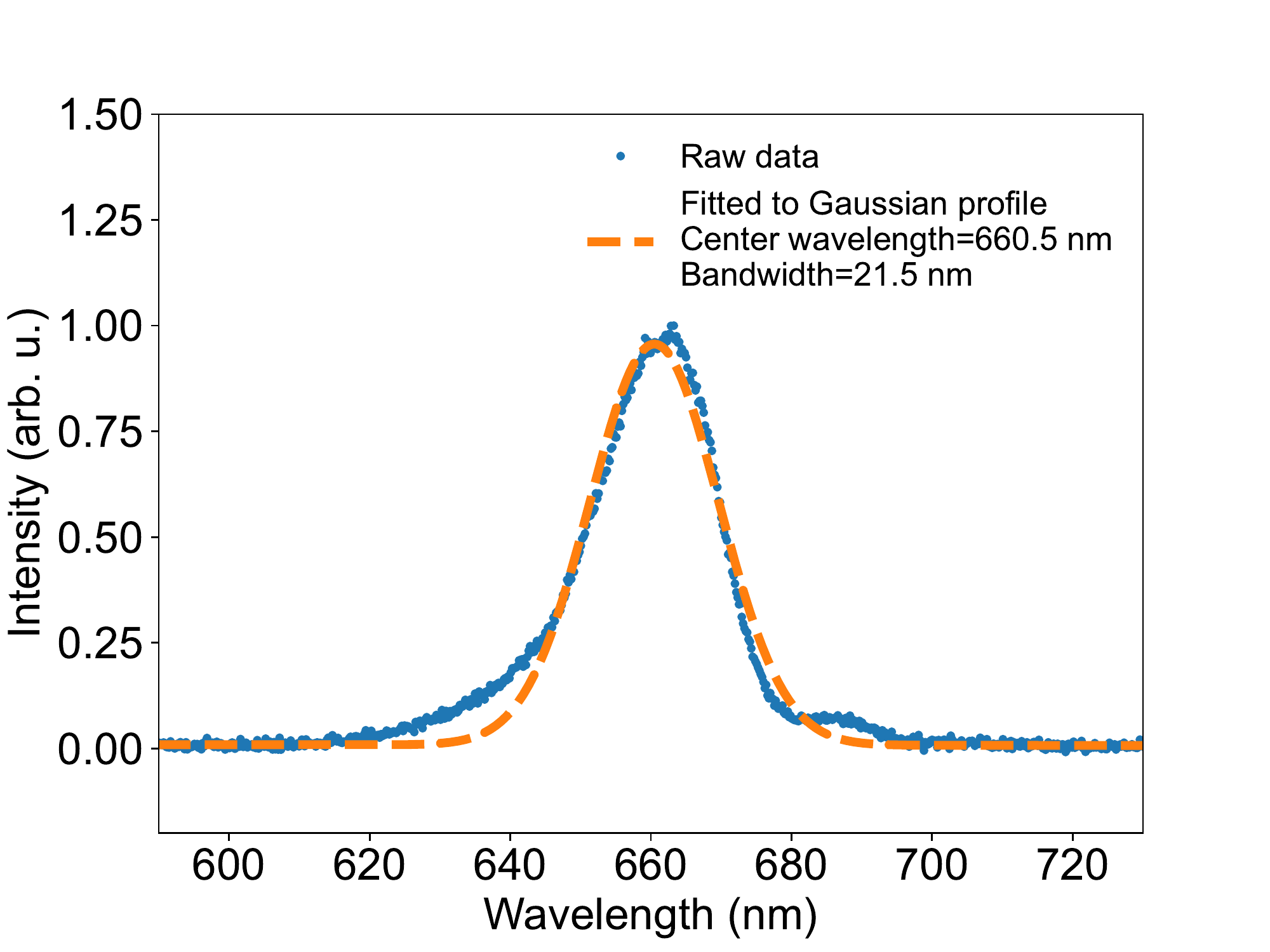}}\hfill
     \subfloat[]{\includegraphics[width = 0.46\textwidth]{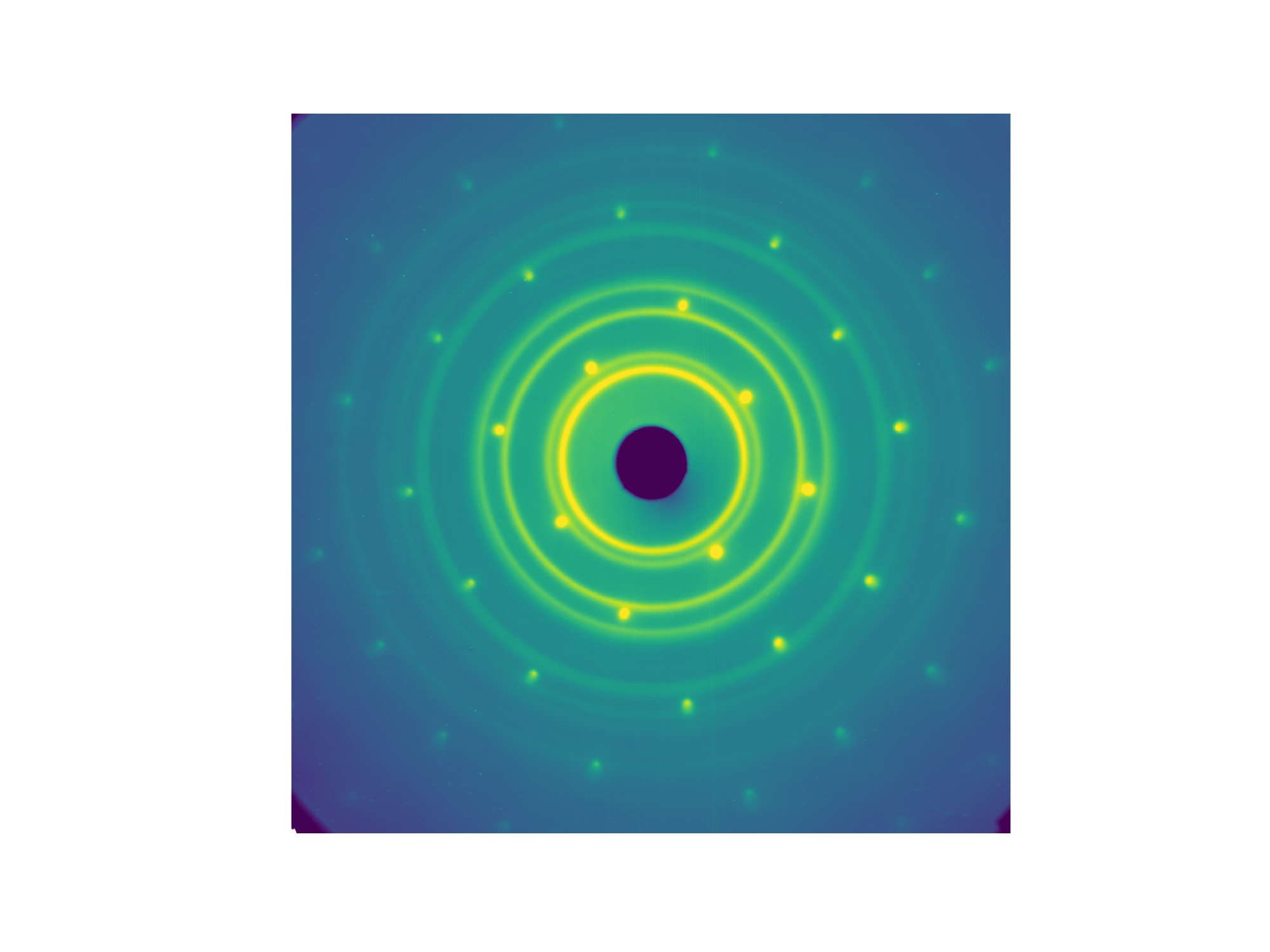}}
     \caption{\textbf{(a)} Schematic of Si-Au heterostructure UED experiment. \textbf{(b)} Pump laser spectrum. \textbf{(c)} Typical diffraction pattern from Si-Au heterostructure.}
     \label{fig:SI_schematic_spectrum_diffraction}
\end{figure}

We carry out the reported experiment in the Megaelectron-volt ultrafast electron diffraction (MeV UED) facility at SLAC National Accelerator Laboratory \cite{weathersby2015mega, shen2018femtosecond}. Fig.\@ \ref{fig:SI_schematic_spectrum_diffraction}(a) shows a schematic of the experimental set up. A Ti:Saphhire regenerative and multi-pass amplifier laser system delivers $800\,\mathrm{nm}$ laser pulses with $60 \,\mathrm{fs}$ full-width-at-half-maximum (FWHM) pulse duration. This laser is split into two parts. The frequency of the first part is tripled to $266 \,\mathrm{nm}$ for electron generation at the photocathode in a radiofrequency (RF) gun, where the electrons are accelerated to $4.2 \,\mathrm{MeV}$ kinetic energy with $<150 \,\mathrm{fs}$ FWHM pulse duration. The other part is used to drive an optical parametric amplifier to generate the pump laser at $650\,\mathrm{nm}$ wavelength. In Fig.\@ \ref{fig:SI_schematic_spectrum_diffraction}(b), we show a measurement of the pump laser spectrum. The electron and pump beams are normally incident onto the sample, where the electron beam is focused to a spot size of $100 \, \mathrm{\mu m}$ full width at half maximum (FWHM) and the pump laser spot size is $300 \, \mathrm{\mu m}$ FWHM. The relative arrival time delay between the pump laser beams and the electron (the delay time) is adjusted by a linear translation stage. The electron diffraction pattern is imaged by a phosphor screen and recorded by an ANDOR iXon Ultra 888 electron multiplying charge-coupled device (EMCCD) camera. A circular through hole in the center of the phosphor allows the passage of the undiffracted electron beam to avoid camera intensity saturation.

\subsection{Data analysis}

\begin{figure}[ht]
    \centering
    \subfloat[]{\includegraphics[height = 5cm]{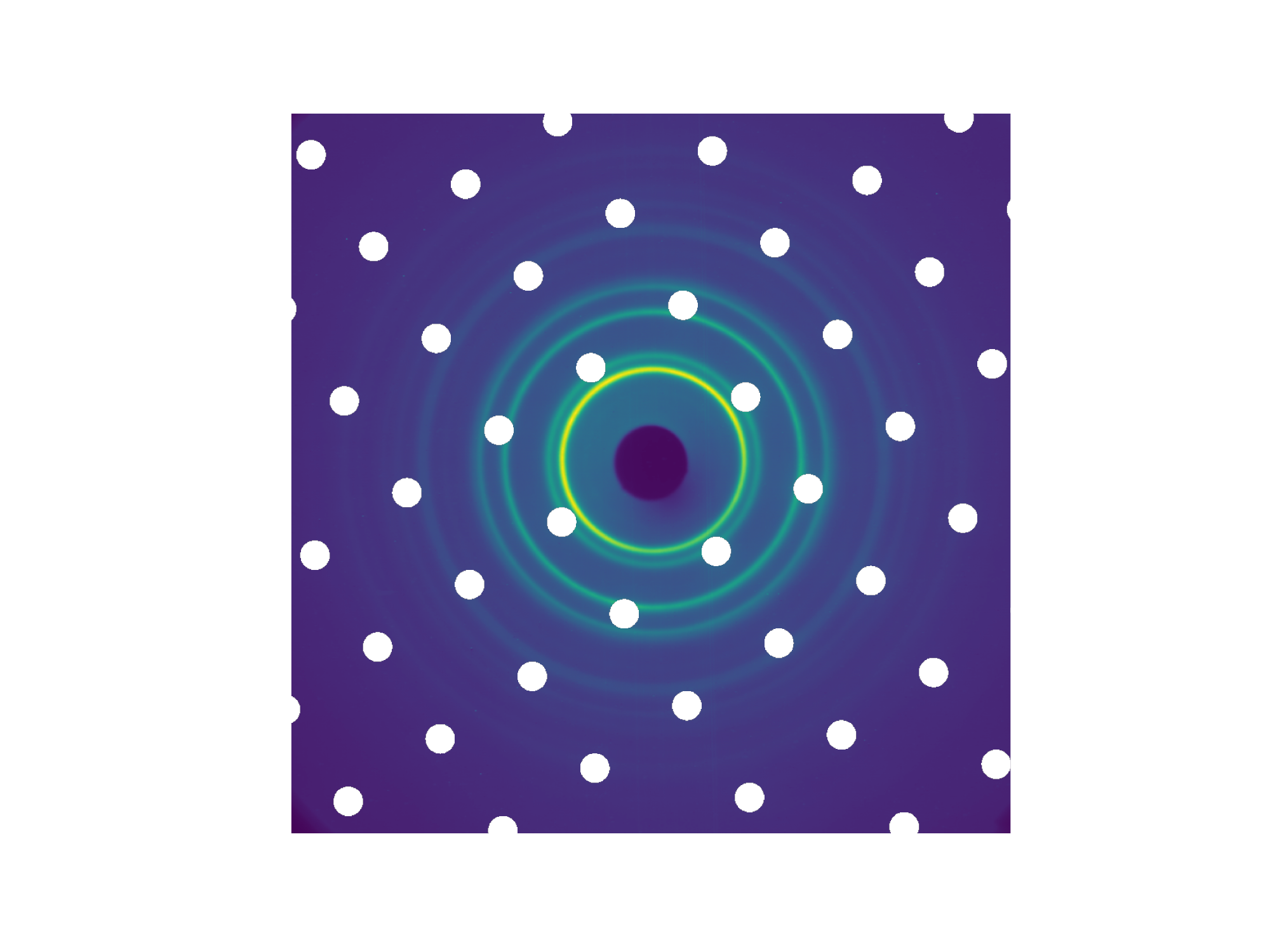}}\hfill
     \subfloat[]{\includegraphics[height = 5cm]{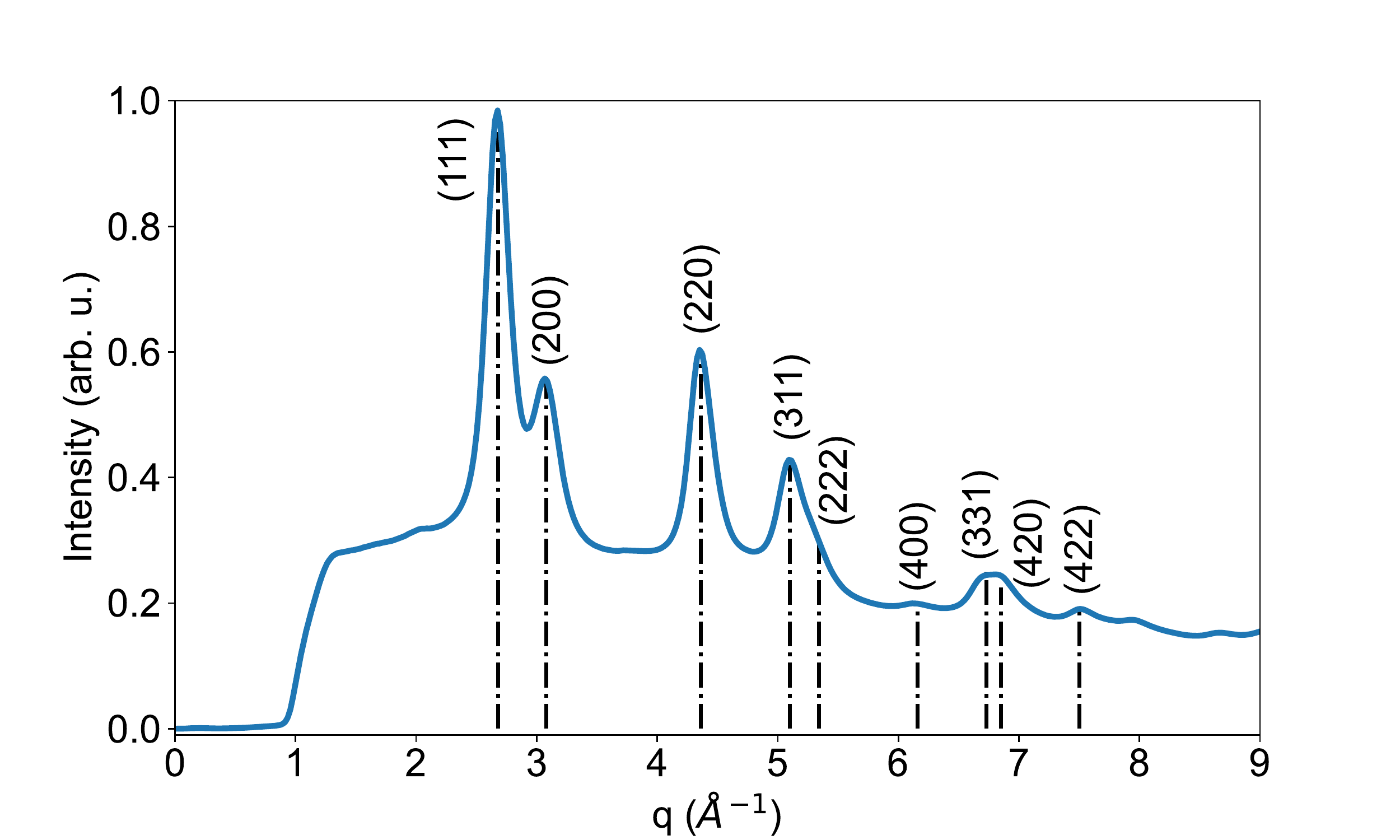}}
    \caption{\textbf{(a)} Diffraction pattern with Si Bragg peaks removed. \textbf{(b)} Azimuthally averaged profile of the diffraction pattern from the panel (a).}
    \label{fig:SI_diffraction_ring_extraction}
\end{figure}

\begin{figure}
     \centering
     \subfloat[]{\includegraphics[height=6cm]{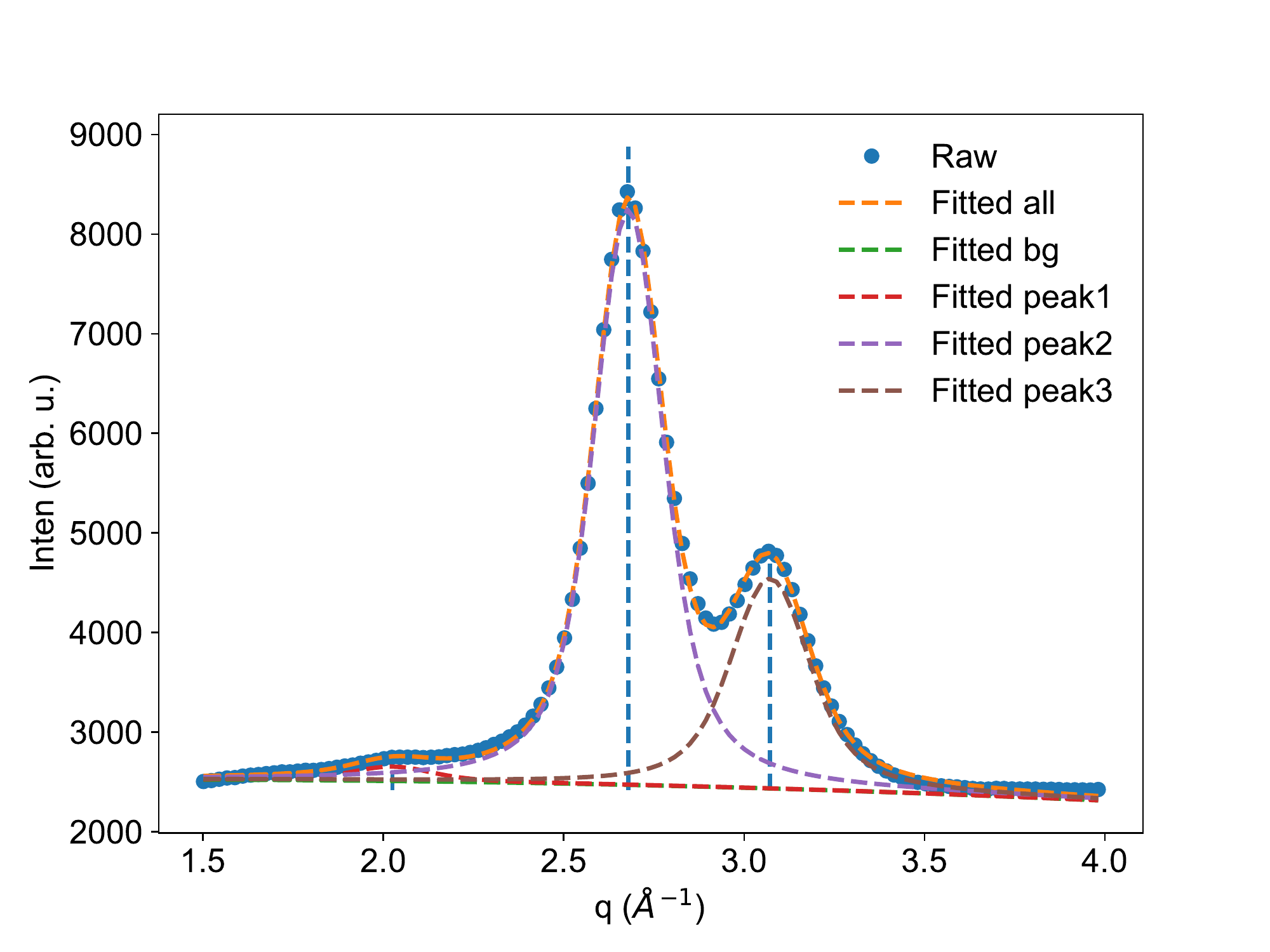}}\hfill
     \subfloat[]{\includegraphics[height=6cm]{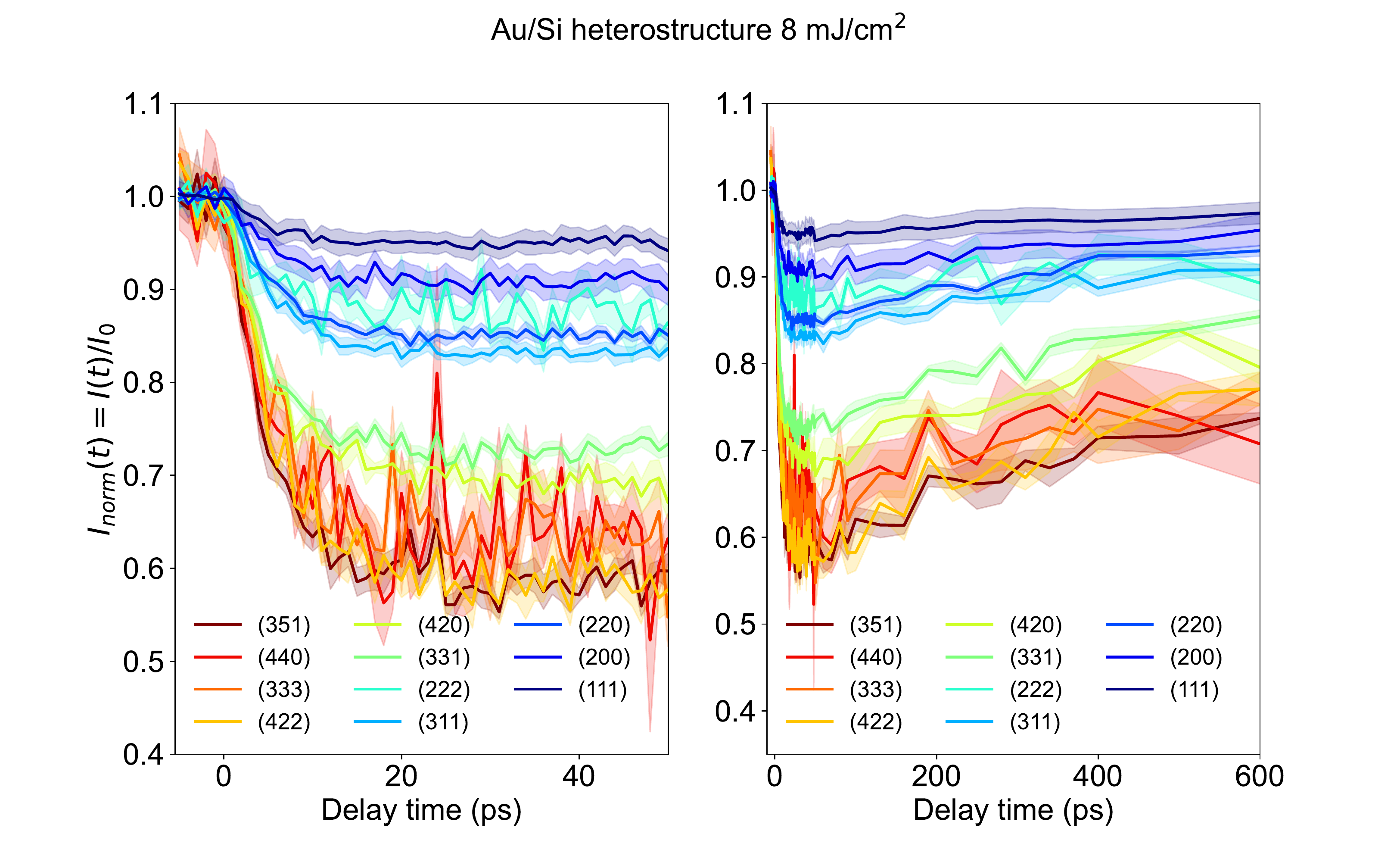}}
     \caption{\textbf{(a)} An example of peak fitting for the radial profile from polycrystalline Au. Voigt profile was implemented for fitting the peaks and a parabolic background was used to for fitting of the smooth background. \textbf{(b)} Normalized intensity $I_\mathrm{norm}(t)$ for selective reflections orders from Au in the Au/Si heterostructure under $8 \mathrm{mJ/cm^{2}}$ laser excitation. While the left plot shows a zoom-in view with delay time within $50\,\mathrm{ps}$, the right plot shows a full delay time range up to $600\,\mathrm{ps}$. Curves with different colors are corresponding to different reflections orders from Au as denoted by the legends. The semi-transparent shaded area shows the standard error of the measurements.}
     \label{fig:SI_peakfitting_Inorm}
\end{figure}

\begin{figure}[hb]
    \centering
    \includegraphics[height=12cm]{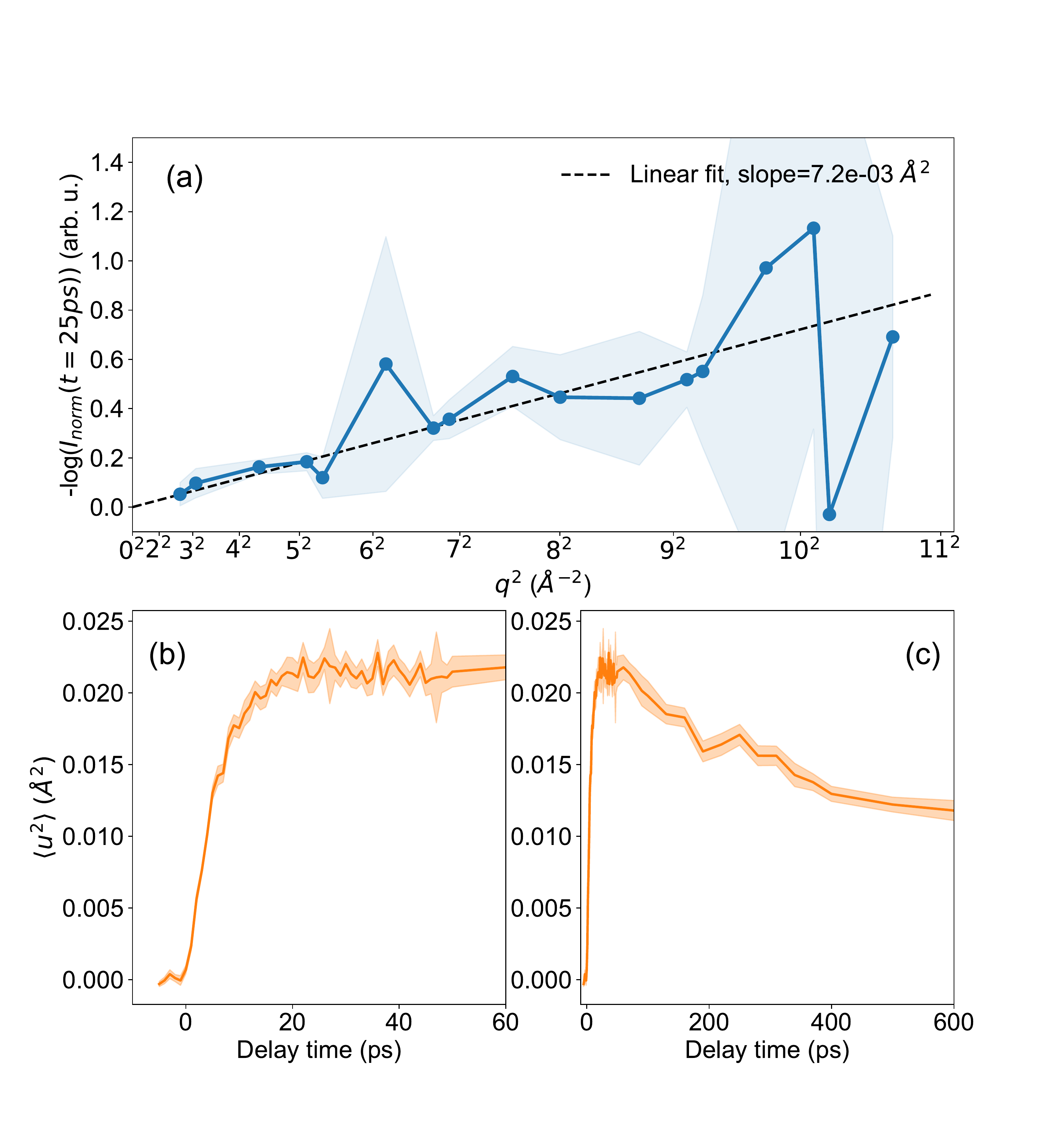}
    \caption{Example of calculation of mean square atomic displacement $\langle u^{2}(t)\rangle$. \textbf{(a)} A representative example of $-\log(I_{\mathrm{norm}}(t,q))$ vs $q^{2}$ for Au from Au/Si heterostructure. A linear fit is performed for estimation of $\langle u^{2}(25\,\mathrm{ps})\rangle$. \textbf{(b)} A zoom-in view of $\langle u^{2}(t)\rangle$ within $60\,\mathrm{ps}$ by repeating the procedure from panel (a). \textbf{(c)} A zoom-out view of $\langle u^{2}(t)\rangle$ within $600\,\mathrm{ps}$ by repeating the procedure from panel (a). The shaded area show the standard errors for the measurements.}
    \label{fig:SI_u2_example}
\end{figure}

We scan the relative arrival time delay between the pump laser beams and the electron over the range of $600 \,\mathrm{ps}$. At each time delay, an electron diffraction image was acquired. The time delay scan is repeated to accumulate data to improve the signal to noise ratio. A typical electron diffraction pattern image from the Au/Si heterostructure captured by the EMCCD camera is shown in Fig.\@ \ref{fig:SI_schematic_spectrum_diffraction}(c). The EMCCD camera background is removed by subtracting the median counts at the four corners of each image. Each image is normalized to its total counts to account for electron beam charge fluctuation during image acquisition. In each image, the Bragg peaks come from single crystal Si, while the Debye-Scherrer rings are from polycrystalline Au. To analyze the diffraction intensity of a Si Bragg peak, two regions of interest (ROIs) of the same size were assigned. One ROI fully covered the Bragg peak, while the other ROI was positioned right next to the Bragg peak and not to cover any ring patterns. The intensity of the Bragg peak is obtained by the difference between the total intensity counts from the two ROIs. For analysis of the Debye-Scherrer ring intensity from polycrystalline Au, the regions of the Si Bragg peaks were first removed from the image, as shown in Fig.\@ \ref{fig:SI_diffraction_ring_extraction}(a). Azimuthal average was then carried out to turn the two-dimensional (2D) patterns into one-dimensional (1D) radial profile, where the 2D Debye-Scherrer rings were reduced to 1D peak profiles correspondingly, as shown in Fig.\@ \ref{fig:SI_diffraction_ring_extraction}(b). To further extract intensity information, peak fittings with Voigt profiles and a parabolic background were applied. An example of peak profiles fitting is shown in Fig.\@ \ref{fig:SI_peakfitting_Inorm}(a). Once the fitted profiles of the peaks were obtained, the total intensities under the peak profiles were evaluated. We repeat the above procedures for diffraction images at all time delays, and obtain the peak intensity as a function of delay time $I(t)$. The time instant when the electron and pump laser beams simultaneously arrive at the sample is set as $t=0$. The steady state intensity $I_{0}$ is obtained from the average of intensities with $t<0$, and the normalized intensity change is calculated by $I_{\mathrm{norm}}(t)=I(t)/I_{0}$. Fig.\@ \ref{fig:SI_peakfitting_Inorm}(b) depicts an example of $I_{\mathrm{norm}}(t)$ for selective reflections from Au. The intensity decrease for a Bragg reflection at momentum transfer $q$ is caused by the mean square atomic displacement $\Delta \langle u^{2}\rangle$ and can be expressed as $I_{\mathrm{norm}}(t, q)=\exp\left(-q^{2}\Delta \langle u^{2}(t)\rangle/3\right)$ (more details are presented in the Section \ref{SI_sec:DW_MSDs}). By fitting a linear function for $-\log(I_{\mathrm{norm}}(t,q))$ versus $q^{2}$, one can extract $\Delta \langle u^{2}(t)\rangle$ as a function of delay time. Fig.\@ \ref{fig:SI_u2_example}(a) shows an example of the linear fitting at $t=25 \,\mathrm{ps}$. Figs.\@ \ref{fig:SI_u2_example}(b) and \ref{fig:SI_u2_example}(c) show $\Delta \langle u^{2}(t)\rangle$ as a function of delay time over $60 \,\mathrm{ps}$ and $600 \,\mathrm{ps}$, respectively.

%% file: anc/SI_boltzmann_transport_equations.tex
\section{Frequency-dependent Boltzmann transport equation}
\label{SI_sec:boltzmann_transport_equations}

\begin{figure}[H]
    \centering
    \includegraphics[width=0.9\linewidth]{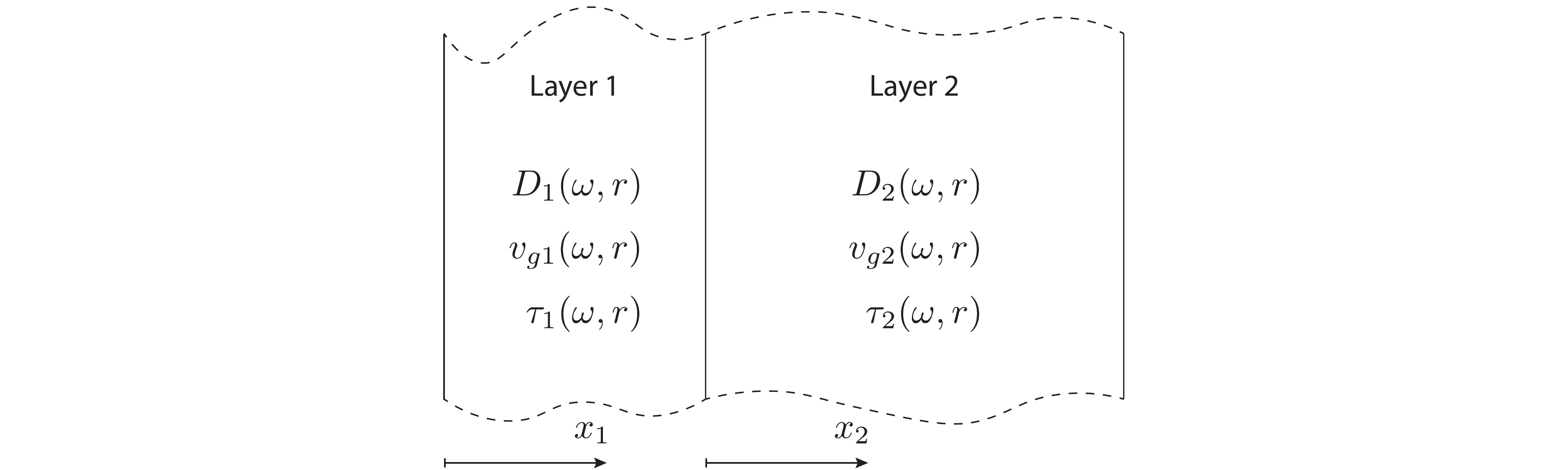}
    \caption{Schematic illustration of the geometry considered in our formulation. A two-layer heterostructure is displayed with spatial coordinates arranged in sequence such that $x_{1}=L_{1}$ and $x_{2}=0$ coincide at the interface, where $L_{n}$ is the thickness of the $n$-th layer and $x_{n}\in[0,L_{n}]$.}
    \label{fig:SI_Layout}
\end{figure}

Given the large ratio of the pump laser spot size of laser pulses ($300\,\mathrm{\mu m}$ FWHM) to the electron probe beam size ($100\,\mathrm{\mu m}$) such that in-plane thermal transport are likely uniform in the probed area, we assume that the cross-plane phonon transport dominates changing diffraction intensities and thus restrict our attention to one-dimensional phonon transport along the thickness direction. In particular, we consider the frequency-dependent Boltzmann transport equation (BTE) with the relaxation time approximation (RTA),
\begin{equation}
    \frac{\partial}{\partial t}g(t,x,\omega,r,\mu)+\mu v(\omega,r)\frac{\partial}{\partial x}g(t,x,\omega,r,\mu)=-\frac{g(t,x,\omega,r,\mu)-g^{\mathrm{eq}}(T_{p}(t,x),\omega,r)}{\tau(\omega,r)},
\end{equation}
where $t$ represents the time and $x$ represents the space coordinate, $v(\omega,r)$ is the frequency- ($\omega$) and polarization- ($r$) dependent group velocity, $\mu=\cos(\varphi)$ is the directional cosine with $\varphi$ being the angle between the phonon propagation direction and the positive $x$-direction, and $T_{p}$ is the pseudo-temperature, as defined later in \eqref{eqn:pseudo_temp_def}. We denote $g(t,x,\omega,r,\mu)=\frac{1}{4\pi}\hbar\omega D(\omega,r) (f(t,x,\omega,r,\mu) - f_{\mathrm{BE}}(\omega, T_{0}))$ as the deviational phonon energy distribution per solid angle, where $f$ is the phonon occupation function, given the density of states (DOS) $D(\omega,r)$. In particular, $f_{\mathrm{BE}}(\omega,T)$ is the Bose-Einstein distribution
\begin{equation}
    f_{\mathrm{BE}}(\omega,T)=\frac{1}{\mathrm{exp}\left(\frac{\hbar\, \omega}{k_{B}T}\right)-1}.
\end{equation} 
The deviational phonon energy distribution in equilibrium states thus reads
\begin{equation}
    g^{\mathrm{eq}}(T(t,x),\omega,r) \coloneqq \frac{1}{4\pi}\hbar\omega D(\omega,r)(f_{\mathrm{BE}}(\omega, T(t,x)) - f_{\mathrm{BE}}(\omega, T_{0}))    
\end{equation} 
where $T_{0}$ is the base temperature that is set to $300\,\mathrm{K}$ throughout this work. For heterostructures, each material layer is governed by a BTE with corresponding phonon properties, such as $v_{n}$, $D_{n}(\omega,r)$, and $\tau_{n}(\omega,r)$ with $n=1,2,\ldots$ being the layer index. $x_{n}\in[0,L_{n}]$ represents the layer-wise spatial coordinate with $L_{n}$ being the layer thickness, as shown in Fig.\@ \ref{fig:SI_Layout}. The phonon energy exchange between layers is described by the interface conditions introduced below.

\subsection{Interface conditions}

We assume diffuse interface between two layers in the considered heterostructure. The boundary conditions for each layer at the interface are determined together with the transmittance and reflectance. Consider the two-layer heterostructure schematically illustrated in Fig.\@ \ref{fig:SI_Layout}. The incoming phonon energy flux in the positive $x_{2}$-direction of layer 2 comprises the flux transmitted from layer 1 and the flux reflected from layer 2, which are represented by the first and the second terms in Eq.\@ \eqref{eqn:ic_full_layer2}, respectively,
\begin{equation}\label{eqn:ic_full_layer2}
    \int_{0}^{1}v_{2} g_{2}(x_{2}=0,\mu)\mu\ \ud \mu = T_{12} \int_{0}^{1}v_{1}g_{1}(x_{1}=L_{1},\mu)\mu\ \ud \mu - R_{21} \int_{-1}^{0}v_{2}g_{2}(x_{2}=0,\mu)\mu\ \ud \mu.
\end{equation}
Similarly, we can obtain the expression for the incoming phonon energy flux from the interface in layer 1 as
\begin{equation}\label{eqn:ic_full_layer1}
    \int_{-1}^{0}v_{1} g_{1}(x_{1}=L_{1},\mu)\mu\ \ud \mu = R_{12} \int_{0}^{1}v_{1}g_{1}(x_{1}=L_{1},\mu)\mu\ \ud \mu - T_{21} \int_{-1}^{0}v_{2}g_{2}(x_{2}=0,\mu)\mu\ \ud \mu.
\end{equation}
For brevity, we omit explicitly writing assumed dependencies of the variables, e.g.\@, $\omega$ and $r$ dependencies in all variables, and the $t$ dependency in $g$. Here we restrict our attention to the case without mode conversion, and denote $T_{12}(\omega,r)$ as the transmittance, or the fraction of total incident energy transmitted from layer 1 to layer 2 across the interface. Correspondingly, the reflection coefficient is determined by $R_{12}(\omega,r)=1-T_{12}(\omega,r)$, or the fraction of energy reflected back to layer 1. One can also determine $R_{21}(\omega,r)=1-T_{21}(\omega,r)$ from $T_{21}(\omega,r)$. 

The principle of detailed balance states that the net energy flux across the interface between two layers vanishes when both sides reach thermal equilibrium at the same temperature $T$, which requires transmittances to satisfy the following condition \cite{swartz1989thermal, minnich2011quasiballistic, hua2017experimental}:
\begin{equation}
    T_{12}(\omega,r)g_{1}^{\mathrm{eq}}(T,\omega,r)v_{1}(\omega,r) = T_{21}(\omega,r)g_{2}^{\mathrm{eq}}(T,\omega,r)v_{2}(\omega,r).
\end{equation}
Given that the equilibrium energy distribution $g^{\mathrm{eq}}(T,\omega,T)$ only depends on $D(\omega,r)$ when $f_{\mathrm{BE}}$ is determined by temperature $T$ and frequency $\omega$, we can use the following alternative:
\begin{equation}
    T_{12}(\omega,r)D_{1}(\omega,r)v_{1}(\omega,r) = T_{21}(\omega,r)D_{2}(\omega,r)v_{2}(\omega,r).
\end{equation}

The above condition allows us to investigate phonon branch resolved transport. However, in this work, we restrict our attention to branch averaged phonon transport. In particular, we make use of the total DOS, i.e., $D_{1}(\omega)=\sum_{r}D_{1}(\omega,r)$, and weighted average phonon properties with branch DOS, for example, the average group velocities reads $v_{1}(\omega)=\sum_{r}D_{1}(\omega,r)v_{1}(\omega,r)/\sum_{r}D_{1}(\omega,r)$. The detailed balance can then be written as
\begin{equation}
    T_{12}(\omega)D_{1}(\omega)v_{1}(\omega) = T_{21}(\omega)D_{2}(\omega)v_{2}(\omega).
\end{equation}


\subsection{Boundary conditions}

Here we follow a similar definition for the material-air boundary conditions as defined in Ref.\@ \citenum{hua2015semi}. Let $\varepsilon_{1}$ and $\varepsilon_{2}$ be the energy loss coefficients of layers 1 and 2 with respect to their adjacent environments; then, the non-black diffuse boundary conditions read
\begin{equation}\label{eqn:au_bc_full}
    \int_{0}^{1}v_{1}g_{1}(x_{1}=0,\mu) \mu\ \ud \mu=\varepsilon_{1} \int_{0}^{1} v_{1} g_{1}^{\mathrm{eq}}(T_{0},\omega) \mu\ \ud \mu - (1-\varepsilon_{1})\int_{-1}^{0}v_{1}g_{1}(x_{1}=0,\mu) \mu\ \ud\mu.
\end{equation}
In particular, we consider that at the boundary $x_{1}=0$, the incoming phonon energies $g_{1}(x_{1}=0,\mu>0)$ distribute evenly and are constant for $\mu>0$, thus $\int_{0}^{1}v_{1}g_{1}(x_{1}=0)\mu\ \ud \mu=\frac{1}{2}v_{1}g_{1}(x_{1}=0,\mu>0)$. A similar condition holds for $\int_{0}^{1}v_{1}g_{1}^{\mathrm{eq}}(T_{0})\mu\ \ud \mu=\frac{1}{2}v_{1}g_{1}^{\mathrm{eq}}(T_{0})$. In addition, the group velocity $v_{g1}(\omega,r)$ is independent of $\mu$ and can be canceled on both sides.
Therefore, Eq.\@ \eqref{eqn:au_bc_full} can be simplified to
\begin{equation}\label{eqn:bc_eps_bdry1}
    g_{1}(x_{1}=0,\mu>0) =\varepsilon_{1} g_{1}^{\mathrm{eq}}(T_{0}) - 2 (1-\varepsilon_{1})\int_{-1}^{0}g_{1}(x_{1}=0,\mu)\mu\ \ud\mu,
\end{equation}
and similarly, we obtain the equation for the other boundary as
\begin{equation}\label{eqn:bc_eps_bdry2}
    g_{2}(x_{2}=L_{2},\mu<0) =\varepsilon_{2} g_{2}^{\mathrm{eq}}(T_{0}) + 2 (1-\varepsilon_{2})\int_{0}^{1}g_{2}(x_{2}=L_{2},\mu)\mu\ \ud\mu.
\end{equation}

\subsection{Initial conditions}
\label{SI_subsec:BTE_ICs}

Proper initial conditions are required for carrying out computations of the BTE. To avoid solving a complex electron-phonon coupled BTE, we choose to focus on the phonon system starting from the moment $t_{0}$ when full relaxation of the electron system is reached, namely when the phonon system no longer exchanges net energy with the electron system. We assume that phonons are subject to the Bose-Einstein distribution of the respective uniform lattice temperatures in each layer at $t=t_{0}$, given by $g(t_{0},x,\omega,r,\mu)=g^{\mathrm{eq}}(T(t_{0},x),\omega,r)=\frac{1}{4\pi}\hbar\omega D(\omega,r)(f_{\mathrm{BE}}(\omega,T(t_{0},x))-f_{\mathrm{BE}}(\omega,T_{0}))$ for each layer, where the initial temperature field $T(t_{0},x)$ is solved from the observed relative diffraction intensities using the Debye-Waller factor (as introduced in Sec.\@ \ref{SI_sec:DW_MSDs}). However, the spatial profile (the $x$-dependence) of the initial temperature field throughout the thickness remains undetermined from relative intensities, so we simply assume a uniform temperature distribution such that $T(t_{0},x)=T(t_{0})$.

\subsection{Other phonon properties}
\label{SI_sec:other_phonon_properties}

Ideally, all phonon properties that appear in the BTE can be treated as model parameters and can be reconstructed given sufficient and proper UED measurements. However, many of these properties are well-studied with \textit{ab initio} calculations, which we can reliably use in order to focus on the remaining properties that are less well-understood, e.g.\@ the transmittances.

First principles-calculated phonon properties from multiple sources are adopted in this work. In particular, the density of states $D(\omega)$, group velocity $v(\omega)$, and relaxation times $\tau(\omega)$ of Al and Si are obtained from Ref.\@ \citenum{hua2016exploring}. The $D(\omega)$ and $v(\omega)$ of Au are calculated with Quantum Espresso \cite{QE-2009, QE-2017} and Phonopy\cite{togo2015first}. We use the $D(\omega)$ and $v(\omega)$ of Al\textsubscript{2}O\textsubscript{3} available from Phonopy examples\cite{PhonopyExample}. The relaxation times of Au and Al\textsubscript{2}O\textsubscript{3} are chosen to be constant, i.e.\@ $\tau_{\mathrm{Au}}\approx 108.17\,\mathrm{ps}$ and $\tau_{\mathrm{Al_{2}O_{3}}}\approx 4.80\,\mathrm{ps}$, such that their bulk thermal conductivities match the order of magnitude of accepted values, $k_{\mathrm{Au}}\approx 310\,\mathrm{Wm^{-1}K^{-1}}$ and $k_{\mathrm{Al_{2}O_{3}}}\approx 30\,\mathrm{Wm^{-1}K^{-1}}$.

%% file: anc/SI_finite_volume_method.tex
\section{Finite volume method for solving BTE}

We employ the finite volume method\cite{eymard2000finite} (FVM) over the space $x$ to evaluate the right-hand-side of the following rearranged BTE,
\begin{equation}
    \frac{\partial}{\partial t}g(t,x,\omega,r,\mu)=-\mu v(\omega,r)\frac{\partial}{\partial x}g(t,x,\omega,r,\mu)-\frac{g(t,x,\omega,r,\mu)-g^{\mathrm{eq}}(\omega,r,T)}{\tau(\omega,r)}.
\end{equation}
The dependencies of the frequency $\omega$, polarization $r$, and direction $\mu$ are considered by solving separate equations at different points $(\omega,r,\mu)$ simultaneously, and thus for brevity we omit these arguments below. For some fixed $(\omega,r,\mu)$ point, we average both sides of the equation over each cell,
\begin{equation}
    \frac{1}{\Delta x}\int_{x_{j}-\frac{\Delta x}{2}}^{x_{j}+\frac{\Delta x}{2}}\frac{\partial}{\partial t}g(t,x)\ \ud x = -\frac{\mu v}{\Delta x}\int_{x_{j}-\frac{\Delta x}{2}}^{x_{j}+\frac{\Delta x}{2}}\frac{\partial }{\partial x}g(t,x)\ \ud x - \frac{1}{\Delta x}\int_{x_{j}-\frac{\Delta x}{2}}^{x_{j}+\frac{\Delta x}{2}}\frac{g(t,x)-g^{\mathrm{eq}}(T(t,x))}{\tau}\ \ud x,
\end{equation}
which can be rewritten with time discretization using the forward Euler method as
\begin{equation} \label{eqn:time_space_discretization}
    \frac{\bar{g}(t_{n+1},x_{j}) - \bar{g}(t_{n},x_{j})}{\Delta t} = -\frac{1}{\Delta x} \left[f(t_{n},x_{j+\frac{1}{2}})-f(t_{n},x_{j-\frac{1}{2}})\right] - \frac{\bar{g}(t_{n},x_{j})-\bar{g}^{\mathrm{eq}}(T(t_{n},x_{j}))}{\tau},
\end{equation}
with $j$ being an integer index of the cell located at $x_{j}$ with boundary surfaces at $x_{j}\pm \Delta x/2$. The cell-averaged values are given by
\begin{equation}
    \bar{g}(t,x_{j})\coloneqq \frac{1}{\Delta x}\int_{x_{j}-\frac{\Delta x}{2}}^{x_{j}+\frac{\Delta x}{2}}g(t,x)\ \ud x,
\end{equation}
while $f(t,x_{j \pm \frac{1}{2} })$ denotes the fluxes at the boundary surfaces defined with the van Leer limiter\cite{dullemond2012lecture}, $\phi(r)\coloneqq \frac{r+|r|}{1+|r|}$. Specifically, for $\mu>0$, when the phonon propagates in the positive $x$-direction, we have
\begin{equation}
\begin{split}
    & f(t,x_{j - \frac{1}{2}}) = \mu v \bar{g}(t, x_{j-1})  + \frac{1}{2} |\mu| v \left(1 - \left|\frac{\mu v \Delta t}{\Delta x}\right|\right) \times \\
    & \qquad \phi(r_{j-\frac{1}{2}})\biggl[\bar{g}(t,x_{j})-\bar{g}(t,x_{j-1})\biggr],\qquad r_{j-\frac{1}{2}}=\frac{\bar{g}(t,x_{j-1})-\bar{g}(t,x_{j-2})}{\bar{g}(t,x_{j})-\bar{g}(t,x_{j-1})}.
\end{split}
\end{equation}
For $\mu<0$, when the phonon propagates in the negative $x$-direction, we have
\begin{equation}
\begin{split}
    & f(t,x_{j - \frac{1}{2}}) = \mu v \bar{g}(t, x_{j})  + \frac{1}{2} |\mu| v \left(1 - \left|\frac{\mu v \Delta t}{\Delta x}\right|\right) \times \\
    & \qquad \phi(r_{j-\frac{1}{2}})\biggl[\bar{g}(t,x_{j})-\bar{g}(t,x_{j-1})\biggr],\qquad r_{j-\frac{1}{2}}=\frac{\bar{g}(t,x_{j+1})-\bar{g}(t,x_{j})}{\bar{g}(t,x_{j})-\bar{g}(t,x_{j-1})}.
\end{split}
\end{equation}
The next time step distribution $\bar{g}(t_{n+1},x_{j})$ can then be computed by rearranging \eqref{eqn:time_space_discretization} as
\begin{equation}
    \bar{g}(t_{n+1},x_{j}) = \bar{g}(t_{n},x_{j}) - \frac{\Delta t}{\Delta x} \left[f(t_{n},x_{j+\frac{1}{2}})-f(t_{n},x_{j-\frac{1}{2}})\right] - \frac{\Delta t}{\tau}\biggl[\bar{g}(t_{n},x_{j})-\bar{g}^{\mathrm{eq}}(T(t_{n},x_{j}))\biggr].
\end{equation}
We implement the above numerical scheme on discretizations listed in Table \ref{tab:discrete_details}.

\begin{figure}[t!]
    \centering
    \includegraphics[width=0.9\linewidth]{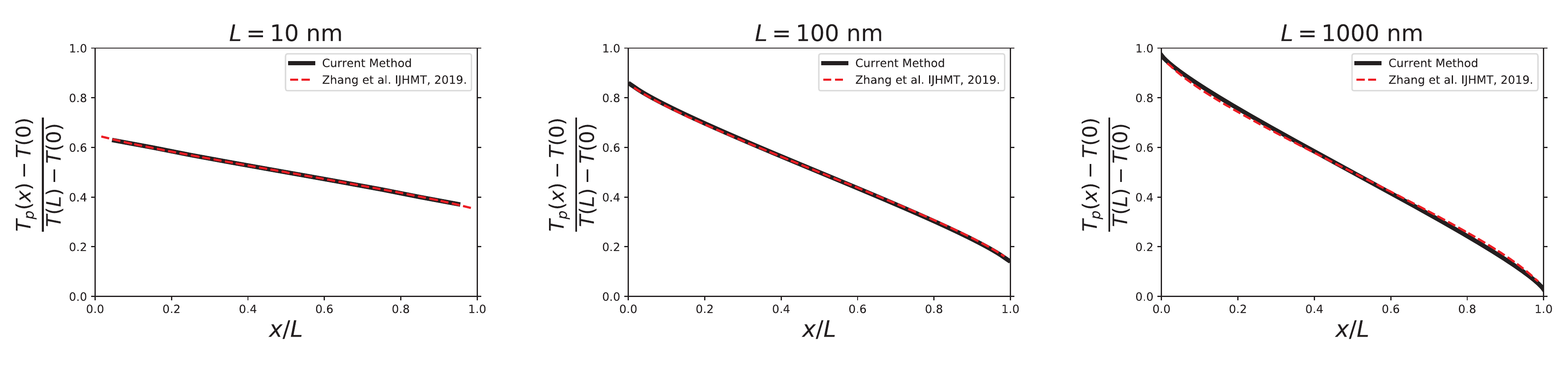}
    \caption{Benchmark solution to verify our implementation of the numerical scheme on a 1D thin-film with fixed boundary conditions and varying thicknesses. The red dashed lines correspond to the results digitally extracted from Fig.\@ 1 in the work by Zhang et al.\cite{zhang2019implicit} using the tool \texttt{WebPlotDigitizer}\cite{Rohatgi2020}.}
    \label{fig:SI_FVM_Benchmark_Linear}
\end{figure}

\begin{table}[!ht]
\centering
\begin{tabular}{|c|c|c|c|}
\hline
$\Delta t$ & $\Delta x$ & $\Delta \omega$ & $\Delta \mu$ \\ 
\hline\hline
$0.1\,\mathrm{ps}$ & $1\,\mathrm{nm}$ & $
0.1,\,0.2\,\mathrm{THz}$ & 0.1 \\ \hline
\end{tabular}
\caption{Details on discretizations. The frequency interval is $0.2\,\mathrm{THz}$ for the Al\textsubscript{2}O\textsubscript{3}/Al heterostructures and $0.1\,\mathrm{THz}$ for the Au/Si heterostructures.}
\label{tab:discrete_details}
\end{table}

\subsection{Verification of numerical implementation}

To verify that the above numerical scheme is correctly implemented, we solve the linearized BTE to compare with existing literature
\begin{equation}
    \frac{\partial}{\partial t}g(t,x,\omega,r,\mu)=-\mu v(\omega,r)\frac{\partial}{\partial x}g(t,x,\omega,r,\mu)-\frac{g(t,x,\omega,r,\mu)}{\tau(\omega,r)}+\frac{1}{4\pi\tau(\omega,r)}C(\omega,r,T_{0})(T-T_{0}),
\end{equation}
where $C(\omega,r,T_{0})$ denotes the specific heat,
\begin{equation}
    C(\omega,r,T_{0})=\hbar\omega D(\omega, r)\frac{\partial f_{\mathrm{BE}}}{\partial T},
\end{equation}
for a thin film with the left wall at $T_{0}+\Delta T$ and right wall at $T_{0}-\Delta T$, given $\Delta T\ll T_{0}$ (here, we select $\Delta T=1\,\mathrm{K}$ and $T_{0}=300\,\mathrm{K}$). The steady-state solution is obtained by iterating forward in time until sufficiently small changes in the temperature profile are observed. We compare our solution with results presented in the work by Zhang et al.\cite{zhang2019implicit} and find excellent agreement, especially for the length scales related to our problem, namely $\sim\mathcal{O}(10^{2}\,\mathrm{nm})$, as shown in Fig.\@ \ref{fig:SI_FVM_Benchmark_Linear}.

\subsection{Solving for lattice temperatures and equilibrium distributions}

\begin{figure}
    \centering
    \includegraphics[width=0.9\linewidth]{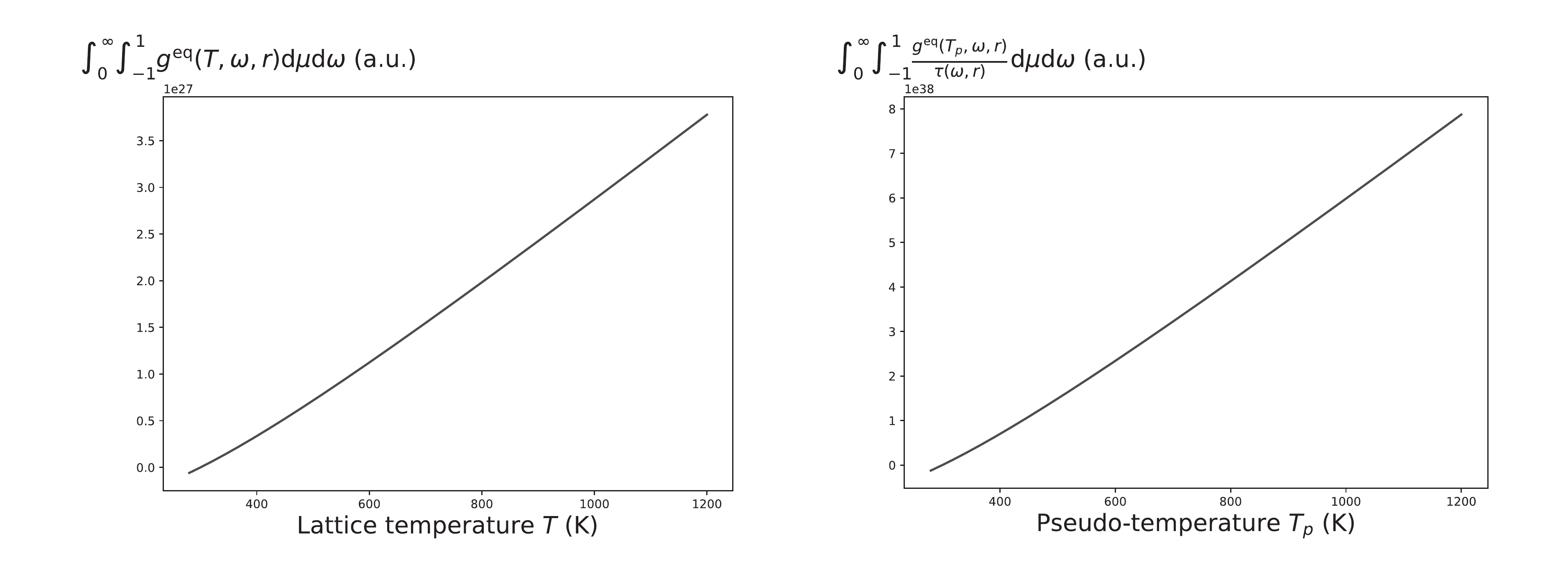}
    \caption{Representative curves for rapid fitting of lattice temperatures $T$ (left) and pseudo-temperatures $T_{p}$ (right) during the forward pass of the BTE model.}
    \label{fig:SI_EngVsTemp}
\end{figure}

When the deviational energy distribution is out of equilibrium, namely $g\neq g^{\mathrm{eq}}$, the concept of temperature is typically not well defined. In this case, we can still obtain an effective measurement by finding a lattice temperature $T(t,x)$ such that the corresponding equilibrium distribution $g^{\mathrm{eq}}(T(t,x),\omega,r)$ has the same total energy as the distribution $g(t,x,\omega,r,\mu)$, which is described by the following equation:
\begin{equation}\label{eqn:lattice_temp_def}
    \sum_{r}\int_{0}^{\infty} \int_{-1}^{1} \left[g(t,x,\omega,r,\mu) - \frac{\hbar\omega}{4\pi} D(\omega,r)\left(f_{\mathrm{BE}}(T(t,x),\omega) - f_{\mathrm{BE}}(T_{0},\omega)\right) \right]\ \ud \mu \, \ud \omega = 0.
\end{equation}
Here we are not restricted to a small temperature rise, and therefore the aforementioned linearization can no longer be used. Given the difficulties to solve the above integral equation, especially during time-marching iterations, we avoid solving it directly and instead prepare a table of points, namely 
\begin{equation}
    (T,\quad \sum_{r}\int_{0}^{\infty} \int_{-1}^{1} \left[\frac{\hbar\omega}{4\pi} D(\omega,r)\left(f_{\mathrm{BE}}(T(t,x),\omega) - f_{\mathrm{BE}}(T_{0},\omega)\right) \right]\ \ud \mu \, \ud \omega),
\end{equation}
so that the transient and local lattice temperatures can be obtained by linear interpolation. 

\begin{figure}[t!]
    \centering
    \includegraphics[width=0.9\linewidth]{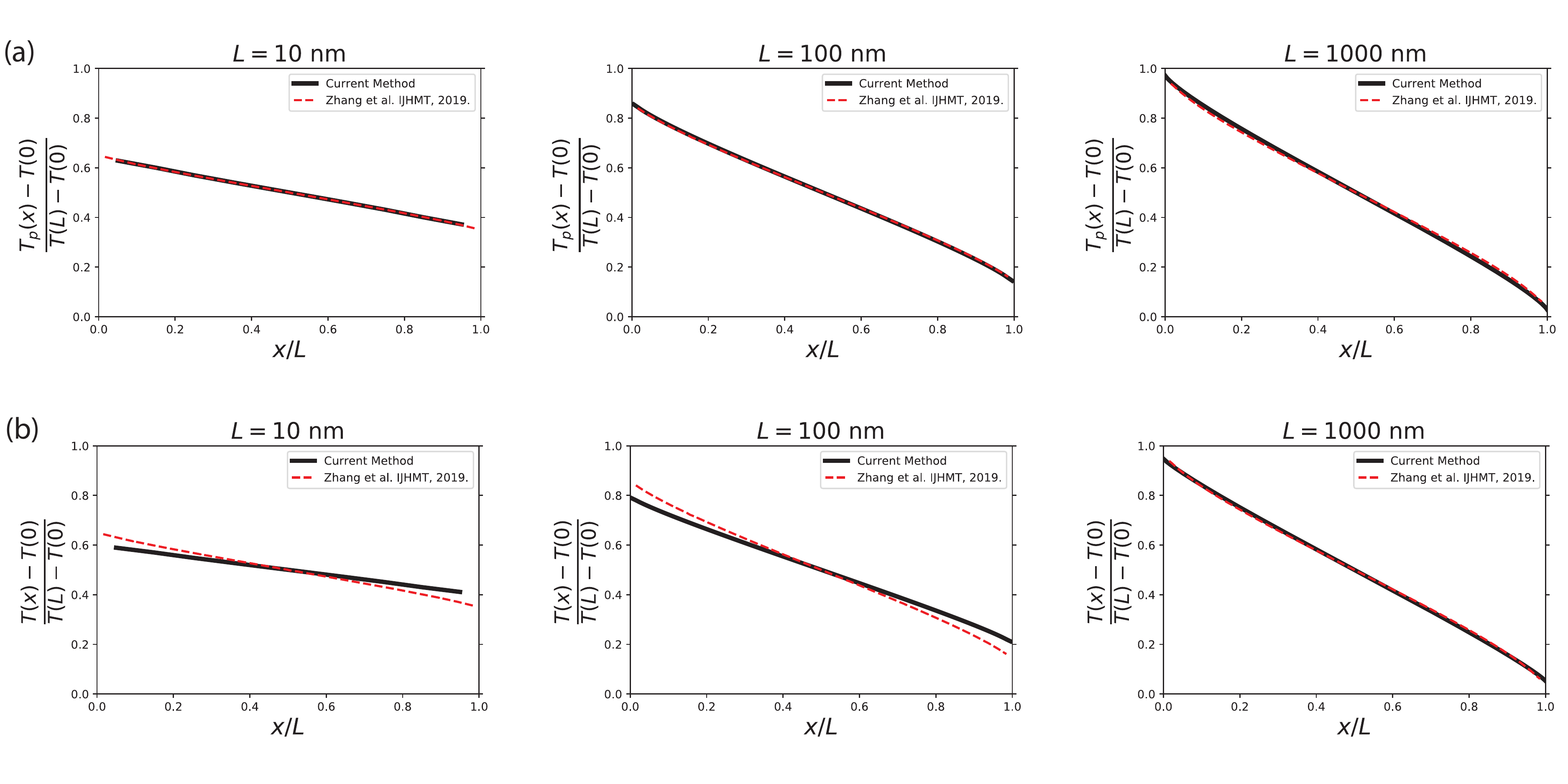}
    \caption{\textbf{(a)} Comparisons of pseudo-temperatures $T_{p}$ between benchmarks and solutions obtained from interpolation. \textbf{(b)} Comparisons of pseudo-temperatures $T_{p}$ from benchmarks and lattice temperatures from the proposed method $T$. Discrepancies are expected due to different definitions of $T_{p}$ and $T$. The temperature resolution of the charts used here are $\delta T=\delta T_{p}=0.001\,\mathrm{K}$ to fit the small temperature rise $\Delta T=1\,\mathrm{K}$. The red dashed lines correspond to the results digitally extracted from Fig.\@ 1 in the work by Zhang et al.\cite{zhang2019implicit} using the tool \texttt{WebPlotDigitizer}\cite{Rohatgi2020}.}
    \label{fig:SI_FVM_Benchmark_NonLinear}
\end{figure}

Notice that at each time step $t_{n}$ and for every spatial point $x_{j}$, we need to calculate the equilibrium energy distribution $\bar{g}^{\mathrm{eq}}(T(t_{n},x_{j}))$ for the collision term. As pointed out by Hao et al.\cite{hao2009frequency}, the pseudo-temperature $T_{p}$ should be introduced to account for transient scattering events, which is obtained by solving
\begin{equation}\label{eqn:pseudo_temp_def}
    \sum_{r}\int_{0}^{\infty} \int_{-1}^{1} \left[\frac{g(t,x,\omega,r,\mu)}{\tau(\omega,r)} - \frac{\hbar\omega}{4\pi\tau(\omega,r)} D(\omega,r)\left(f_{\mathrm{BE}}(T_{p}(t,x),\omega) - f_{\mathrm{BE}}(T_{0},\omega)\right) \right]\ \ud \mu \, \ud \omega = 0.
\end{equation}
Similarly, we also prepare a function chart that includes a list of pre-computed points,
\begin{equation}
    (T_{p},\quad \sum_{r}\int_{0}^{\infty} \int_{-1}^{1} \left[\frac{\hbar\omega}{4\pi\tau(\omega,r)} D(\omega,r)\left(f_{\mathrm{BE}}(T_{p}(t,x),\omega) - f_{\mathrm{BE}}(T_{0},\omega)\right) \right]\ \ud \mu \, \ud \omega),
\end{equation}
to obtain real-time solutions of $T_{p}$ for arbitrary distributions $g$. In Fig.\@ \ref{fig:SI_EngVsTemp}, we show two representative tables that are used to interpolate $T$ and $T_{p}$. By solving the same benchmark problem as in Fig.\@ \ref{fig:SI_FVM_Benchmark_Linear}, we show that the proposed method can reach great accuracy by comparison to $T_{p}$ in Fig.\@ \ref{fig:SI_FVM_Benchmark_NonLinear}(a). However, due to differences in definitions, the lattice temperatures $T$ are typically different from the pseudo-temperatures $T_{p}$, as shown in Fig.\@ \ref{fig:SI_FVM_Benchmark_NonLinear}(b). To obtain the results displayed in this work (excluding those shown in Fig.\@ \ref{fig:SI_FVM_Benchmark_NonLinear}), the temperature interval of both charts is set to $\delta T=\delta T_{p}=0.01\,\mathrm{K}$.

%% file: anc/SI_debye_waller.tex
\section{Mean-squared displacements (MSD\lowercase{s}) of thermal vibrations from relative electron diffraction intensities} \label{SI_sec:DW_MSDs}

According to the Debye-Waller effect, the electron diffraction intensity can be attenuated by thermal vibration of atoms by a factor of $\exp\left(-\frac{1}{3}q^{2}\langle u^{2}\rangle\right)$, and thus the relative intensity reads
\begin{equation}
    I(t,q)=\frac{e^{-\frac{1}{3}q^{2}\langle u^{2}(t)\rangle}}{e^{-\frac{1}{3}q^{2}\langle u_{0}^{2}\rangle}}=e^{-\frac{1}{3}q^{2}\Delta \langle u^{2}(t)\rangle},
\end{equation}
where $q=|\mathbf{q}|$ is the norm of the wavevector, and $\langle u^{2}\rangle$ denotes the mean-squared displacement. $\langle u_{0}^{2}\rangle$ denotes the MSD corresponding to the base temperature $T_{0}$, namely before the action of the laser pump, and $\Delta \langle u^{2}\rangle=\langle u^{2}(t)\rangle-\langle u_{0}^{2}\rangle$ represents the difference after laser heating. For crystals with only one atomic species, this quantity is given by the relation \cite{als2011elements} 
\begin{equation}\label{eqn:MSD_ss}
    \langle u^{2}\rangle(t,x) = \frac{1}{N m}\sum_{r}\int_{0}^{\infty}\frac{1}{\omega^{2}}\left(g(t,x,\omega,r)+\frac{\hbar \omega}{2}D(\omega)\right)\ \ud \omega,
\end{equation}
where we have integrated over the solid angle to obtain
\begin{equation}
    g(t,x,\omega,r) = 2\pi \int_{-1}^{1}g(t,x,\omega,r,\mu)\ \ud \mu.
\end{equation}
The corresponding expression of $\langle u^{2}(t,x)\rangle$ for multiple species is derived in Section \ref{sec:MSD_MultiSpec}. Since the measured intensities $I(t,q)$ are not spatially resolved, to compare with computed $\langle u^{2}(t,x)\rangle$, the extracted MSDs are obtained by averaging over $x$ through
\begin{equation}
    \Delta \langle u^{2}(t)\rangle = \frac{1}{L}\int_{0}^{L}\Delta \langle u^{2}(t,x)\rangle\ud x.
\end{equation}

\subsection{Mean-squared displacement (MSD) for general crystals}
\label{sec:MSD_MultiSpec}

Following the relation between the single mode phonon energy and the kinetic energy for simple substances, for crystals with multiple atomic species in the unit cell, we have
\begin{equation}
\begin{split}
    \frac{1}{2}N_{1}m_{1}\omega^{2}\langle u_{1}^{2}\rangle + \frac{1}{2}N_{2}m_{2}\omega^{2}\langle u_{2}^{2}\rangle & = \frac{1}{2}\left(g(\omega)+\frac{\hbar \omega}{2}D(\omega)\right) \\
    & = \frac{1}{2}p_{1}(\omega)\left(g(\omega)+\frac{\hbar \omega}{2}D(\omega)\right) + \frac{1}{2}p_{2}(\omega)\left(g(\omega)+\frac{\hbar \omega}{2}D(\omega)\right),
\end{split}
\end{equation}
where $p_{j}(\omega)=D_{j}(\omega)/D(\omega)$ denotes the ratio of the partial DOS of $j$th-element to the total DOS, and for any frequency points with $D(\omega)>0$, we have $\sum_{j}p_{j}(\omega)=1$. Thus the mode-averaged MSD is given by
\begin{equation}
    \langle u_{j}^{2}\rangle = \frac{1}{N_{j}m_{j}} \int_{0}^{\infty}\frac{p_{j}(\omega)}{\omega^{2}}\left(g(\omega)+\frac{\hbar \omega}{2}D(\omega)\right)\ud \omega.
\end{equation}
However, the element-wise MSD is not accessible from UED measurement, and only effective MSDs averaged over all atoms are available. Since the unit cell structure factor has a complex form given the number of species $N_{\mathrm{sp}}$,
\begin{equation}
    F^{\mathrm{u.c.}}(\mathbf{q})=\sum_{j}^{N_{\mathrm{sp}}}\sum_{n=1}^{N_{j}}f_{j}(\mathbf{q})e^{-M_{j}}e^{i\mathbf{q}\cdot\mathbf{r}_{n}},
\end{equation}
which results in $\mathbf{q}$-dependent contributions to the relative intensity of each element $\langle u_{j}^{2}\rangle$. Therefore, it is rather difficult to obtain a relation for general crystals in a similar way for simple substances \eqref{eqn:MSD_ss}. Here we adopt the following simplification to approximate the effective MSD:
\begin{equation}
    \langle u^{2}\rangle=\sum_{j}^{N_{\mathrm{sp}}}\frac{N_{j}}{N_{\mathrm{tot}}}\langle u_{j}^{2}\rangle, \qquad \text{with } N_{\mathrm{tot}}=\sum_{j}^{N_{\mathrm{sp}}} N_{j};
\end{equation}
namely, we assign fixed weights according to number of atoms of each element.

%% file: anc/SI_optimization_method.tex
\section{Reconstructions of phonon properties}
\label{SI_sec:optimization_method}

\subsection{Adjoint state method}

This section aims to provide a brief derivation of the adjoint equations solved in this work, adapted from the process presented in Ref. \citenum{chris2020lecture}. A more detailed introduction of the adjoint-state method can be found in other works \cite{cao2003adjoint}. Suppose that we have extracted from electron diffraction intensities the MSDs as a function of delay time $\Delta\langle u^{2}\rangle^{\mathrm{target}}$, and simulated intensities $\Delta\langle u^{2}\rangle$ based on model parameters $\theta$. The total loss is therefore an integral of temporal loss $\hat{l}(g(\theta),\langle u^{2}\rangle^{\mathrm{target}})=l(\langle u^{2}\rangle, \langle u^{2}\rangle^{\mathrm{target}})$ spanning the duration of the measurements,
\begin{equation}
    \hat{L}(g,\theta) = \int_{t_{0}}^{t_{f}}\hat{l}(g(\theta),\langle u^{2}\rangle^{\mathrm{target}})\, \ud t + \int_{t_{0}}^{t_{f}} \lambda(t)\left(\frac{\partial}{\partial t}g(\theta) - f(g(\theta),\theta)\right) \,\ud t,
\end{equation}
where we have introduced a Lagrange multiplier $\lambda(t)$ to incorporate constraints on $g$ imposed by the Boltzmann transport equations. It should be noted that we reused the symbol $f(.)$ to denote a function with parameters $\theta$ that takes as input the current phonon energy distribution $g$ and outputs the changing rate $\partial g/\partial t$, namely $\partial g/\partial t=f(g;\theta)$. The gradients of the loss with respect to parameters are
\begin{equation}
    \frac{\partial \hat{L}}{\partial \theta}=\int_{t_{0}}^{t_{f}}\left(\frac{\partial \hat{l}}{\partial \theta} + \frac{\partial \hat{l}}{\partial g}\frac{\partial g}{\partial \theta}\right) \, \ud t + \int_{t_{0}}^{t_{f}} \lambda(t)\left(\frac{\partial}{\partial t}\frac{\partial g}{\partial \theta} - \frac{\partial f}{\partial g}\frac{\partial g}{\partial \theta} - \frac{\partial f}{\partial \theta}\right) \, \ud t.
\end{equation}
We can perform integration by parts on the second term and rearrange the expression to obtain
\begin{equation}
\begin{split}
    \frac{\partial \hat{L}}{\partial \theta} & = \int_{t_{0}}^{t_{f}}\left(\frac{\partial \hat{l}}{\partial \theta} + \frac{\partial \hat{l}}{\partial g}\frac{\partial g}{\partial \theta}\right) \, \ud t + \left.\lambda\frac{\partial g}{\partial \theta}\right|_{t=0}^{t_{f}} - \int_{t_{0}}^{t_{f}} \left(\frac{\partial \lambda}{\partial t}\frac{\partial g}{\partial \theta} + \lambda\frac{\partial f}{\partial g}\frac{\partial g}{\partial \theta} + \lambda\frac{\partial f}{\partial \theta}\right) \, \ud t \\
    & = \int_{t_{0}}^{t_{f}}\left(\frac{\partial \hat{l}}{\partial \theta} - \lambda\frac{\partial f}{\partial \theta}\right) \, \ud t + \lambda(t_{f})\frac{\partial g(t_{f})}{\partial \theta} - \int_{t_{0}}^{t_{f}} \left(\frac{\partial \lambda}{\partial t} + \lambda\frac{\partial f}{\partial g} - \frac{\partial \hat{l}}{\partial g} \right) \frac{\partial g}{\partial \theta} \, \ud t,
\end{split}
\end{equation}
where we have used the fact that the initial condition $g(t_{0})$ is independent of model parameters $\theta$, and thus $\left.\frac{\partial g}{\partial \theta}\right|_{t=t_{0}} = 0$. It is possible to avoid evaluating any of $\partial g/\partial \theta$ by satisfying the following conditions:
\begin{equation} \label{eqn:adj_g_ode}
    \frac{\partial \lambda}{\partial t} = - \lambda\frac{\partial f}{\partial g} + \frac{\partial \hat{l}}{\partial g},\qquad \lambda(t_{f})=0,
\end{equation}
where the adjoint $\lambda(t)$ should be solved backward in time from $t=t_{f}$ to $t_{0}$. We are left with $\frac{\partial \hat{L}}{\partial \theta}=\int_{t_{0}}^{t_{f}}\left(\frac{\partial \hat{l}}{\partial \theta} - \lambda\frac{\partial f}{\partial \theta}\right) \, \ud t$. Notice that the temporal loss $\hat{l}$ does not directly depend on the parameters $\theta$, and thus $\partial \hat{l}/\partial \theta=0$. (However $\ud \hat{l}/\ud \theta$ is not zero given that $g$ depends on $\theta$.) With an appropriate initial condition $\eta(t_{f})=0$, the $\eta(t^{\prime})$ for some intermediate moment $t_{0}\le t^{\prime}\le t_{f}$ thus denotes the gradient $\partial \hat{L}/\partial \theta$ evaluated based on measurements collected over the time range $[t^{\prime},t_{f}]$. We can evaluate the overall gradient $\frac{\partial \hat{L}}{\partial \theta}=\eta(t_{0})$ (given all measurements in $[t_{0},t_{f}]$) by solving the following equation backward from $t_{f}$ to $t_{0}$:
\begin{equation} \label{eqn:adj_theta_ode}
    \frac{\partial \eta}{\partial t}=-\lambda\frac{\partial f}{\partial \theta},\qquad \eta(t_{f})=0.
\end{equation}
In practice, we do not have continuous but only discrete experiment electron diffraction intensities, represented by the set of MSDs $\{t_{n}, \langle u^{2}(t_{n})\rangle^{\mathrm{target}}\}_{n=1}^{N}$. Therefore, when integrating \eqref{eqn:adj_g_ode}, the direct gradient term $\partial \hat{l}/\partial g$ is always zero between data points (i.e.\@, when $t_{n}<t<t_{n+1}$ for $n=1,\ldots,(N-1)$) except when exactly $t=t_{n}$; there, it becomes nonzero and contributes to the backward evolution of $\lambda(t)$. Here, we adopt mean-absolute loss to measure the reconstruction quality as defined by
\begin{equation}
    L = \frac{1}{N}\sum_{i=1}^{N}l_{n},\qquad l_{n}=|\langle u^{2}(t_{n})\rangle-\langle u^{2}(t_{n})\rangle^{\mathrm{target}}|.
\end{equation}
All necessary derivatives appearing in \eqref{eqn:adj_g_ode} and \eqref{eqn:adj_theta_ode} are computed by automatic differentiation (AD) implemented in PyTorch\cite{NEURIPS2019_9015}, and the implementation of the adjoint-state method with AD is adapted from the Ref.\@ \citenum{chen_neural_2018} and GitHub repositories\cite{Chen2018GitHub, Mikhail2019GitHub} with certain modifications, including the reuse of pre-computed (in the forward pass) $g(t,x,\omega,r,\mu)$ during the backward computation to guarantee numerical stability.

\subsection{Initialization and reconstruction of phonon properties}
\label{SI_subsec:init_params}

In general, the proposed framework can be applied to multi-layer heterostructures and any phonon properties of interest as long as the partial derivatives appearing in \eqref{eqn:adj_g_ode} and \eqref{eqn:adj_theta_ode} can be computed. In this work, we restrict our discussion to the two-layer heterostructure (Fig.\@ \ref{fig:SI_Layout}) and restrict the properties to energy loss coefficients $\varepsilon_{1/2}\in [0,1]$, and transmittances $T_{12/21}(\omega)\in[0,1]$. Due to the wide range of possible relaxation times, it is relatively difficult to reconstruct them, and some prior knowledge is required for successful reconstruction, in particular their potential functional forms. We leave more details about relaxation times below in Section \ref{SI_subsec:recon_T_tau}. The phonon properties being reconstructed are initialized by randomly drawing from uniform distributions over their admissible ranges; specifically, we have $\varepsilon^{(0)}, \mathcal{T}^{(0)}(\omega)\sim\mathcal{U}[0,1]$. The coefficients related to relaxation times of Au are randomly initialized from $\tau_{n}\sim\mathcal{U}[0,216]\,\mathrm{ps}$ such that the mean value corresponds to the average relaxation tim as defined in Section \ref{SI_sec:other_phonon_properties}.

As introduced above, we obtain the simulated MSDs by forward calculation of the BTE-based model, which is followed by backward propagation to produce the gradients $\partial L/\partial \theta$. Then updates can be made with gradient-based methods, e.g., gradient descent $\theta\to \theta-\alpha \frac{\partial L}{\partial \theta}$ with some learning rate $\alpha$, and its variants. Here, we select the Adam optimizer \cite{kingma2014adam} with $\alpha=0.01$ to update energy loss coefficients and transmittances and $\alpha=2.5$ for the relaxation times related parameters, where the learning rate is determined from random search and may be further optimized.

\begin{figure}
    \centering
    \includegraphics[width=0.6\linewidth]{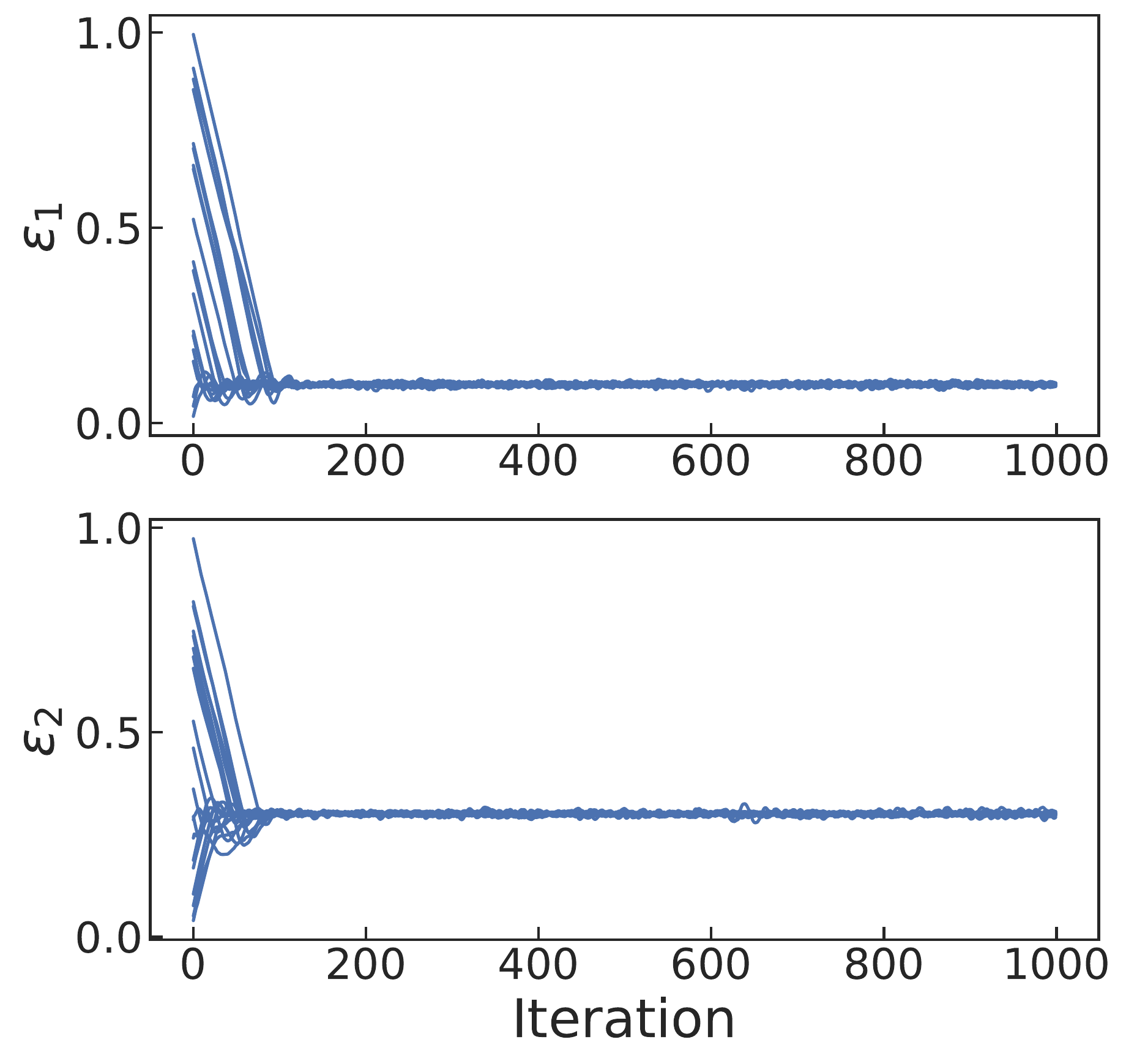}
    \caption{$\varepsilon$ versus iteration, where each line represents a randomly initialized parameter and its update history. Their mean values and standard deviations are then visualized in the main text. This selected example corresponds to the first subfigure of the Fig.\@ 2(b).}
    \label{fig:SI_Epsilon_vs_Iteration}
\end{figure}

\begin{figure}
    \centering
    \includegraphics[width=0.9\linewidth]{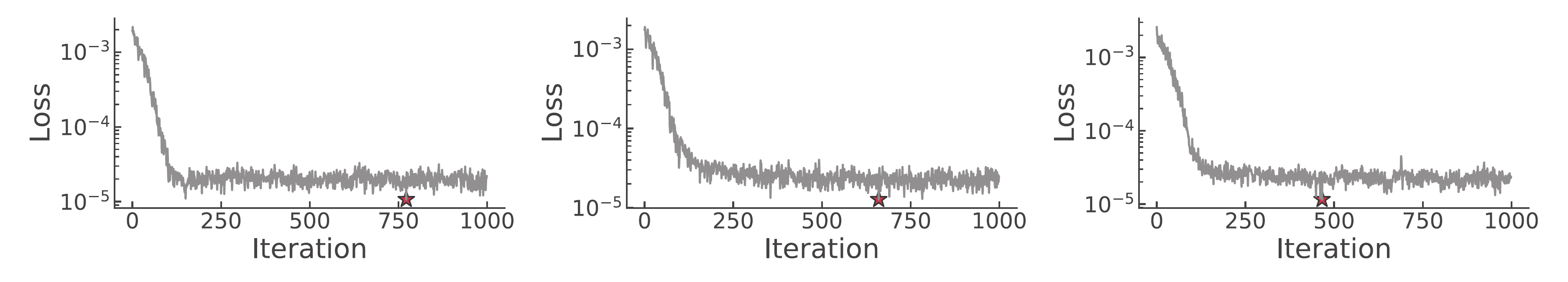}
    \caption{Loss histories for the cases demonstrated in Fig.\@ 2(c) of the main text in the same order. The red star indicates the iteration with the lowest error, and the corresponding reconstructions are displayed in Fig.\@ 2(c).}
    \label{fig:SI_LossHist_Fig2}
\end{figure}

In practical reconstructions, we always independently draw and optimize multiple sets of parameters $\theta$ to exclude the influence of different initializations, as exemplified by $\varepsilon_{1}$ and $\varepsilon_{2}$ in Fig.\@ \ref{fig:SI_Epsilon_vs_Iteration}. For each iteration, a randomly-drawn initial condition from the prepared pool (training samples, as listed in Table \ref{SI_tab:initial_conditions_Al2O3_Al} and \ref{SI_tab:initial_conditions_Au_Si}) is used, and all parameter sets are updated simultaneously. We choose a sufficiently large maximum iteration number and consider the parameters resulting in the lowest error as our reconstructed results. For example, the loss histories of results presented in Fig.\@ 2(c) in the main text are visualized in Fig.\@ \ref{fig:SI_LossHist_Fig2}.

\begin{table}[]
    \centering
    \begin{tabular}{cccccccc}
        \hline
        Material & $L$ (nm) & \multicolumn{5}{c}{$T(t_{0})$ (K)} \\
        \hline
        \hline
        Al\textsubscript{2}O\textsubscript{3} & 5  & 400  & 600  & 800  & 800  & 400 \\
        Al    & 35 & 400  & 600  & 800  & 400  & 800\\
        \hline
    \end{tabular}
    \caption{Initial conditions used in reconstructions of $\theta=\{\varepsilon_{1/2},T_{12/21}(\omega)\}$ with the rest being fixed at true values.}
    \label{SI_tab:initial_conditions_Al2O3_Al}
\end{table}


%% file: anc/SI_more_recon_tasks.tex
\section{More reconstruction tasks}
\label{SI_sec:more_recon_tasks}

\subsection{Representative mean absolute gradients of parameters}

\begin{figure}[h]
    \centering
    \includegraphics[width=0.6\linewidth]{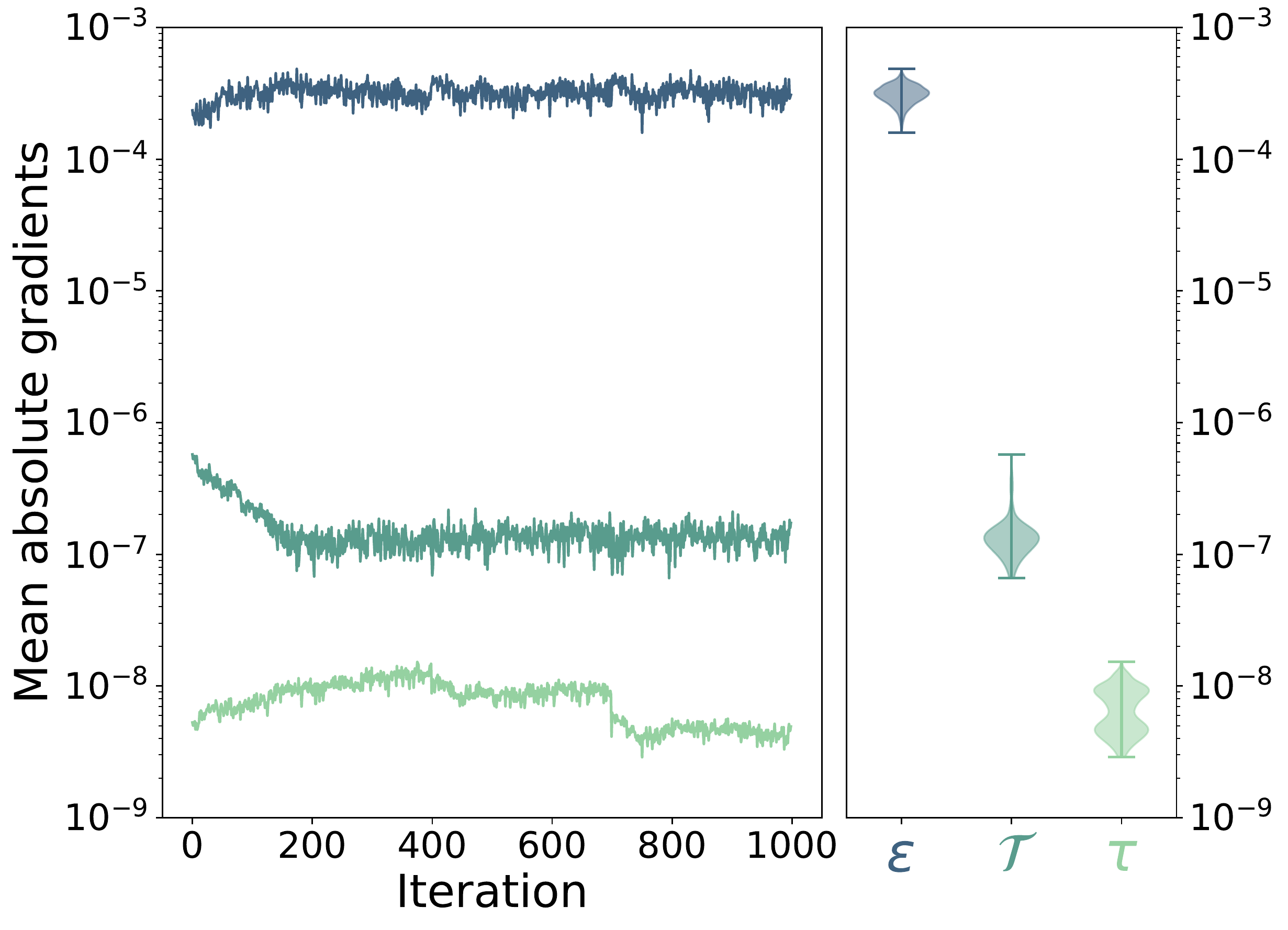}
    \caption{A representative example for the mean absolute magnitude of gradients of the loss with respect to different parameters, with each color corresponding to a parameter as indicated in the right panel.}
    \label{fig:SI_gradients}
\end{figure}

We compute the average absolute gradients of the loss with respect to each phonon property as a function of training epochs as sensitivity analysis in the left panel of Fig.\@ \ref{fig:SI_gradients}, where the right panel displays the distribution over the entire training process. The correspondences between curves in the left panel and properties can be found through the horizontal axis of the right panel. The magnitude of the gradient $|\partial L/\partial \varepsilon|$ prevails over the entire training process. The $|\partial L/\partial \mathcal{T}|$ remain under $10^{-6}$, which is linked to their higher uncertainties and the larger number of iterations necessary for convergence shown in Fig.\@ 2(c). We note the even lower magnitudes of gradients related to relaxation times, which suggests the highest reconstruction difficulties and the least robustness should be expected if three properties are being reconstructed simultaneously.

\subsection{Reconstructions of transmittances only}

\begin{figure}[h]
    \centering
    \includegraphics[width=0.9\linewidth]{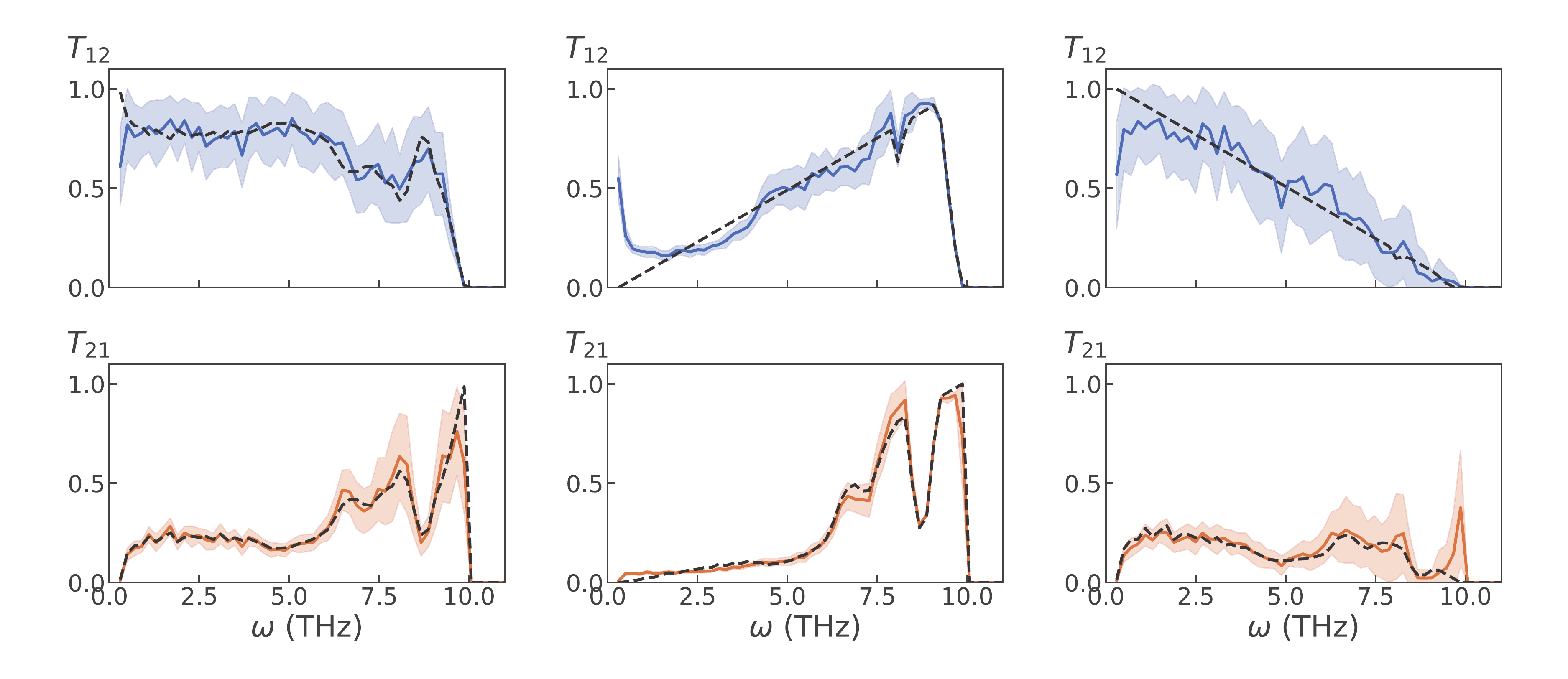}
    \caption{Reconstructions of transmittances with all other parameters fixed at their true values for the Al\textsubscript{2}O\textsubscript{3}/Al heterostructure. The black dash lines represent the ground truths.}
    \label{fig:SI_Recon_TR_Only}
\end{figure}

Here we consider comparatively easier reconstruction tasks, assuming the only unknown is the transmittances, and all other phonon properties can be obtained either from measurements or computation with high confidence. Thus, all other parameters except the transmittances are fixed at their true values. The reconstructed transmittances are shown in Fig.\@ \ref{fig:SI_Recon_TR_Only}, where remarkable improvement can be found when compared with the results of mixed reconstructions like Fig.\@ 2 in the main text.

\subsection{Simultaneous reconstructions of transmittances and relaxation times}
\label{SI_subsec:recon_T_tau}

\begin{figure}
    \centering
    \includegraphics[width=0.9\linewidth]{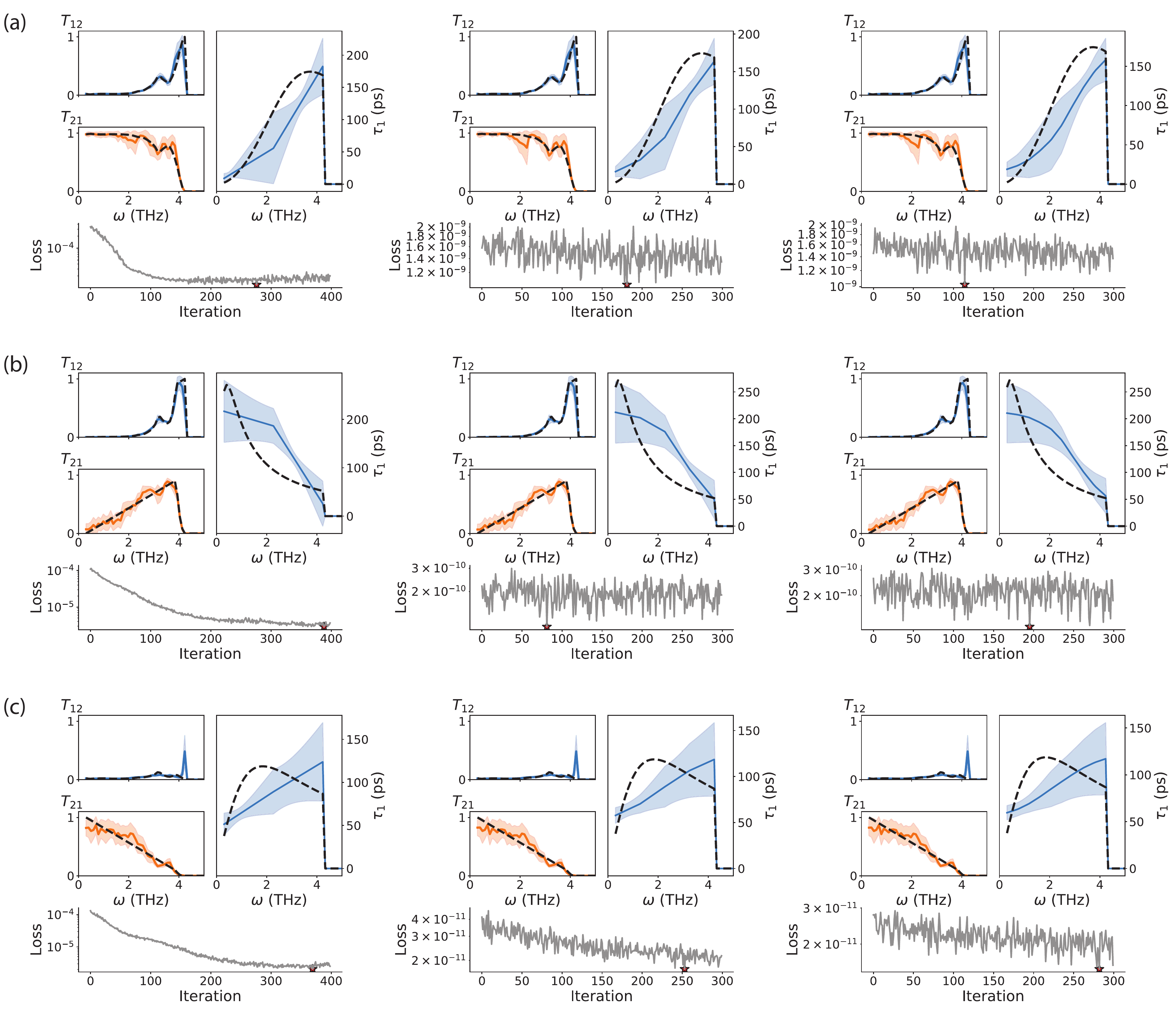}
    \caption{Simultaneous reconstructions of transmittances and the relaxation times $\tau_{\mathrm{Au}}$ for the Au/Si heterostructure, each row of panels corresponds to a distinct combination of transmittances and relaxation times. The red star under each panel indicate the epoch of transmittances and relaxation times being displayed.}
    \label{fig:SI_Tau_fitting}
\end{figure}

Here we demonstrate that our framework can be extended to reconstruct frequency-dependent relaxation times by adopting a multi-stage reconstruction strategy to progressively capture details of relaxation times $\tau$ (inspired by Ref.\@ \citenum{forghani2019phonon}), as shown in Fig.\@ \ref{fig:SI_Tau_fitting}. We use piece-wise linear basis functions $\phi_{n}(\omega)$ to form a lower-dimensional representation with the format $\tau(\omega)=\sum_{n=1}^{N}\tau_{n}\phi_{n}(\omega)$, where $\tau_{n}$'s become the new set of trainable parameters. This allow us to fit fewer coordinates $\tau_{n}$ instead of the full profile and thus reduces reconstruction difficulties in the beginning, but maintains the capability to capture details as the number of basis functions increases. In particular, we start from 3 basis functions and perform the optimization for 400 epochs (left panels of Fig.\@ \ref{fig:SI_Tau_fitting}) with the mean absolute loss $L=\sum_{n=1}^{N} \frac{1}{N} |\langle u^{2}(t_{n})\rangle-\langle u^{2}(t_{n})\rangle^{\mathrm{target}}|$. Then we interpolate the best reconstructed profile (correspond to the smallest loss marked by the red star) to 5 basis and conduct 300 epochs of optimization (middle panels of Fig.\@ \ref{fig:SI_Tau_fitting}), the same process is repeated for 10 basis with 300 epochs (right panels of Fig.\@ \ref{fig:SI_Tau_fitting}). The optimizations on 5 and 10 basis are done with mean absolute loss, $L=\sum_{n=1}^{N} \frac{1}{N} |\langle u^{2}(t_{n})\rangle-\langle u^{2}(t_{n})\rangle^{\mathrm{target}}|^{2}$, to achieve better convergence. Six sets of initial conditions are used as listed in Table \ref{SI_tab:initial_conditions_Au_Si}. 

We comment that the reconstruction performances of relaxation times may not be as robust as other phonon properties, as can be seen from panel Fig.\@ \ref{fig:SI_Tau_fitting}(c). The difficulties can be largely attributed to small differences induced by $\tau$ in the length and time scales considered in this work, which are reflected by the gradients $\partial L/\partial \tau_{n}$ shown in Fig.\@ 3(a). Further improvements on more reliably restoring relaxation times are open for future studies.

\begin{table}[]
    \centering
    \begin{tabular}{ccccccccc}
        \hline
        Material & $L$ (nm) & \multicolumn{5}{c}{$T(t_{0})$ (K)} \\
        \hline
        \hline
        Au & 5  & 400  & 495.6  & 600  & 700  & 800 & 900 \\
        Si & 35 & 500  & 610.3  & 700  & 800  & 900 & 1000 \\
        \hline
    \end{tabular}
    \caption{Initial conditions used in reconstructions of $\theta=\{T_{12/21}(\omega),\tau_{1}(\omega)\}$ with the rest being fixed at true values as shown in the Fig.\@ 3 of the main text.}
    \label{SI_tab:initial_conditions_Au_Si}
\end{table}

%% file: anc/SI_AuSi_numexp.tex
\section{More on analyzing experimental UED measurements of the A\lowercase{u}/S\lowercase{i} heterostructure}

\subsection{Replacing the outlier point with the interpolation}

\begin{figure}[h]
    \centering
    \includegraphics[width=0.5\linewidth]{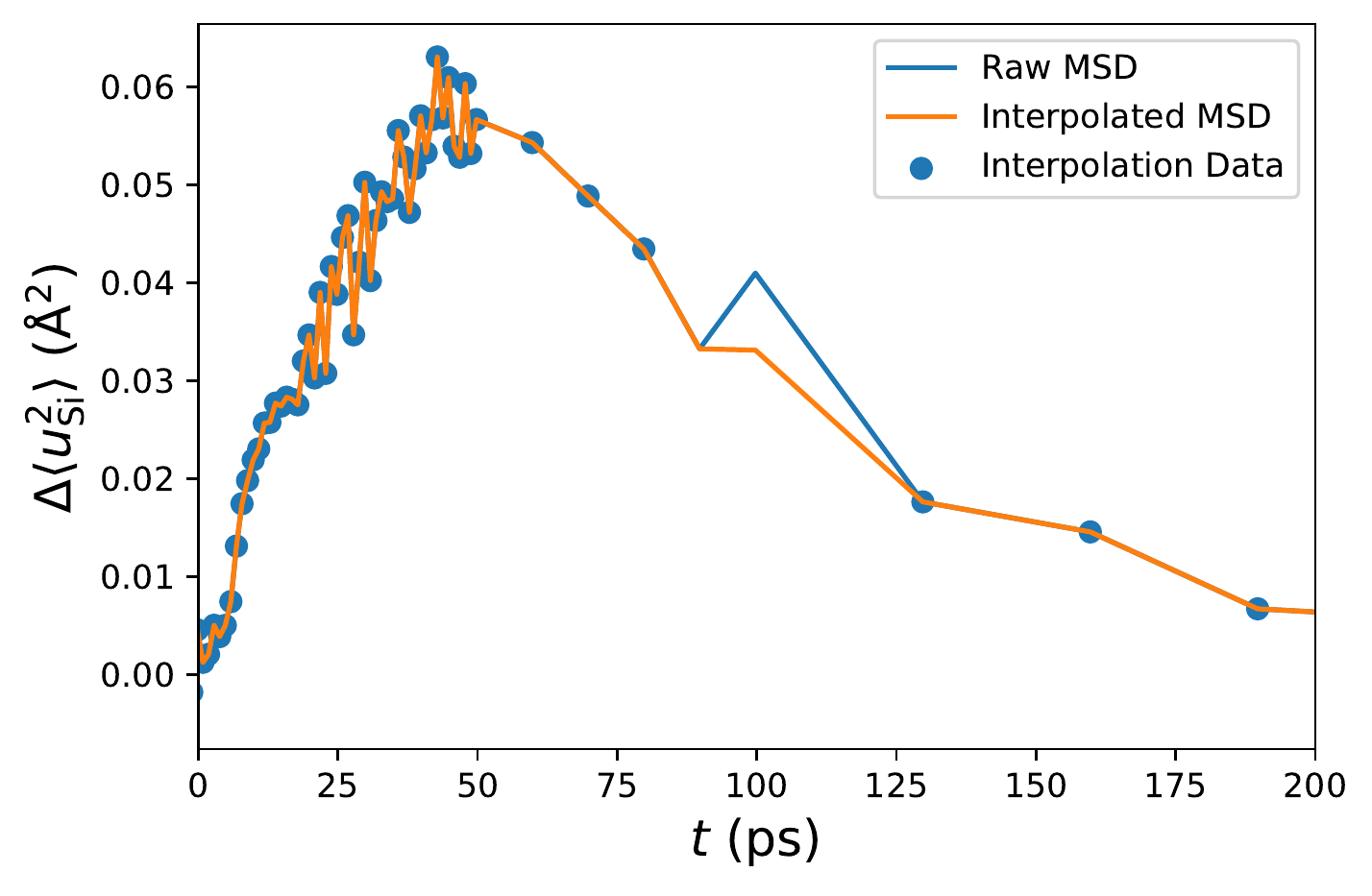}
    \caption{Illustration of the pre-process made on the experimentally-extracted MSDs of the Si $\Delta \langle u_{\mathrm{Si}}^{2}\rangle$.}
    \label{fig:SI_interp_SiExpData}
\end{figure}

We noticed an obvious outlier in the experimentally extracted MSDs of the Si $\langle u_{\mathrm{Si}}^{2}\rangle$ at around $t\approx 99.75\,\mathrm{ps}$. To minimize influences induced by this outlier, we linearly interpolate its value based on MSDs at $t\approx 79.76\,\mathrm{ps}$ and $t\approx 129.73\,\mathrm{ps}$, as illustrated in Fig.\@ \ref{fig:SI_interp_SiExpData}.

\subsection{Determination of initial conditions}

In the presented experimental UED measurements of the Au/Si heterostructure, it is not straightforward to identify $t_{0}$ from MSDs, as $\langle u_{\mathrm{Au}}^{2}\rangle$ and $\langle u_{\mathrm{Si}}^{2}\rangle$ may not reach their respective maximum values at the same delay time. In addition, the presence of noise makes it difficult to locate the exact delay time that corresponds to the maximum MSD for each layer. Remember that we defined $t_{0}$ as the moment when the electron and phonon systems stop exchanging energy and share the same lattice temperature; roughly speaking, the phonon energies will be maximized at $t_{0}$. Thus, a rigorous approach to determine $t_{0}$ would be to compute the total phonon energy of the entire heterostructure system as a function of delay time, $E(t)$, and then identify $t_{0}=\arg\max_{t}E(t)$. Here we adopt a simplified approach by noticing the existence of plateau in $\langle u_{\mathrm{Au}}^{2}(t)\rangle$ and thus identify the starting time $t_{0}$ as the one with largest $\langle u_{\mathrm{Si}}^{2}(t)\rangle$. This particular choice results in $t_{0}\approx 42.8\,\mathrm{ps}$, $T_{\mathrm{Au}}(t_{0})\approx 495.6\,\mathrm{K}$, and $T_{\mathrm{Si}}(t_{0})\approx 610.3\,\mathrm{K}$.

\subsection{Numerical demonstrations using the same settings}
\label{SI_subsec:numexp_AuSi}

Due to the lack of ``ground truth'' in experimental measurements, it is relatively difficult to evaluate the quality of our reconstruction in Fig.\@ 4 of the main text. However, we can ask our model to fit from synthetic MSDs produced from exactly the same settings as those from the experimental measurement, including initial temperatures and sampling time points, but with prescribed energy loss coefficients and transmittances being ``ground truths''.

\begin{figure}
    \centering
    \includegraphics[width=0.95\linewidth]{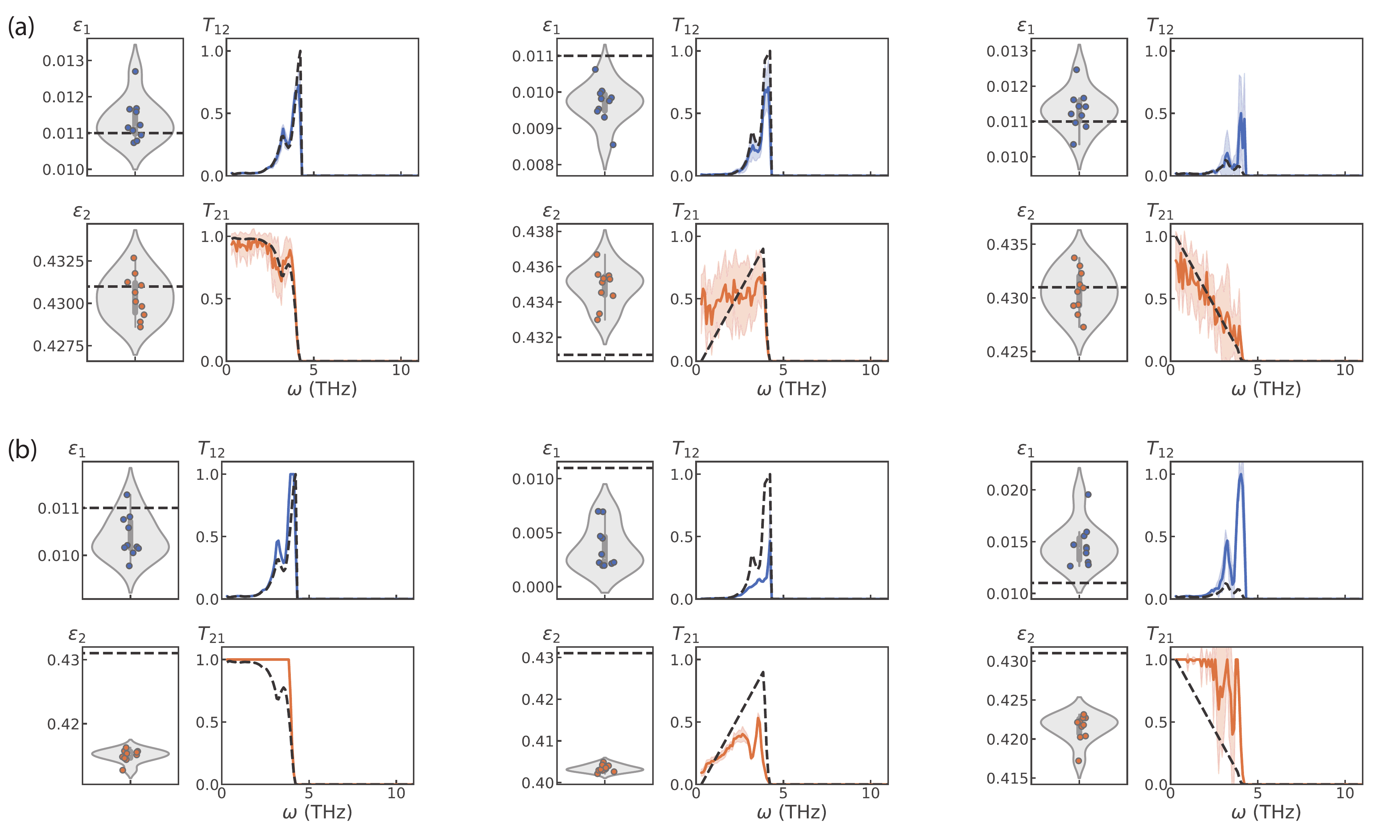}
    \caption{\textbf{(a)} Comparisons between reconstructed and true phonon properties from clean synthetic MSDs. \textbf{(b)} Comparisons between reconstructed and true phonon properties from noisy synthetic MSDs with $\delta=0.05$. The statistics presented here comes from 10 set of randomly initialized parameters.}
    \label{fig:SI_AuSi_NumExp}
\end{figure}

We conducted numerical experiments for three sets of parameters without and with $\delta=0.05$ artificial noise, as shown in Fig.\@ \ref{fig:SI_AuSi_NumExp}(a) and (b), respectively. We can see satisfactory agreement in the reconstructions of both energy loss coefficients and transmittances from clean synthetic MSDs. However, our approach seems to perform significantly worse, especially in terms of transmittances, in the presence of artificial noise. One important reason for the degraded performance in the presence of noise is that there is only one training sample available, compared with multiple training samples as listed in Table \ref{SI_tab:initial_conditions_Al2O3_Al} and \ref{SI_tab:initial_conditions_Au_Si}. It should also be noted that the simulated noisy MSDs might represent a different scenario than real experimental measurements, as each sampling time point represents one realization in a uniform distribution, while each experimental data point is actually averaged from several measurements. Unfortunately, we have no quantitative measure of biases of experimental means, thus we cannot simulate them in our numerical experiments.

Though the proposed framework cannot necessarily provide us with reliable reconstructions due to limited measurements in this example, it can be utilized to guide experimental design, for instance, by informing the minimum amount of information required from experiment for robust reconstructions, i.e., the number of different laser energy inputs (corresponding to initial lattice temperatures), sampling time intervals, and etc.

\subsection{Different phenomenological descriptions of energy loss}\label{SI_sec:bulk_energy_loss}

In addition to the boundary conditions Eq.\@ \eqref{eqn:bc_eps_bdry1} and \eqref{eqn:bc_eps_bdry2} where the energy losses are parameterized by $\varepsilon$'s (below we call them $\varepsilon^{\mathrm{Bdry}}$ since they are defined on boundaries), we also consider the formulation that describes energy loss through bulk of each layer
\begin{equation}
    \frac{\partial g}{\partial t}+v\frac{\partial g}{\partial x}=-\frac{g-g^{\mathrm{eq}}(T_{p})}{\tau} - \varepsilon^{\mathrm{Bulk}}(g-g^{\mathrm{eq}}(T_{0})),
\end{equation}
where $T_{p}$ is the pseudo-temperature corresponding to the current energy distribution $g$, while $T_{0}$ is the reference (room) temperature.

\begin{figure}
    \centering
    \includegraphics[height=0.4\linewidth]{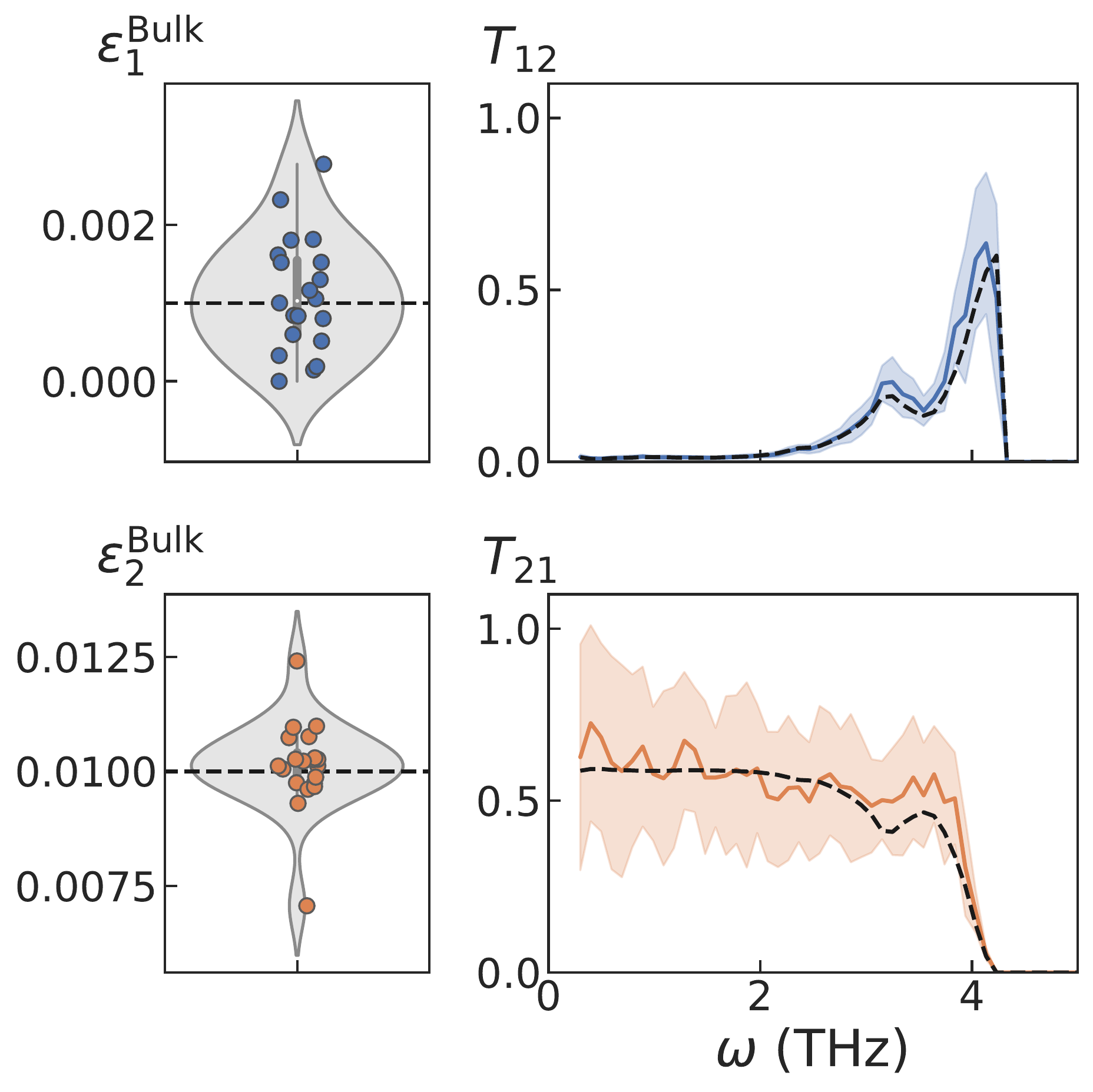}
    \hspace{25pt}
    \includegraphics[height=0.4\linewidth]{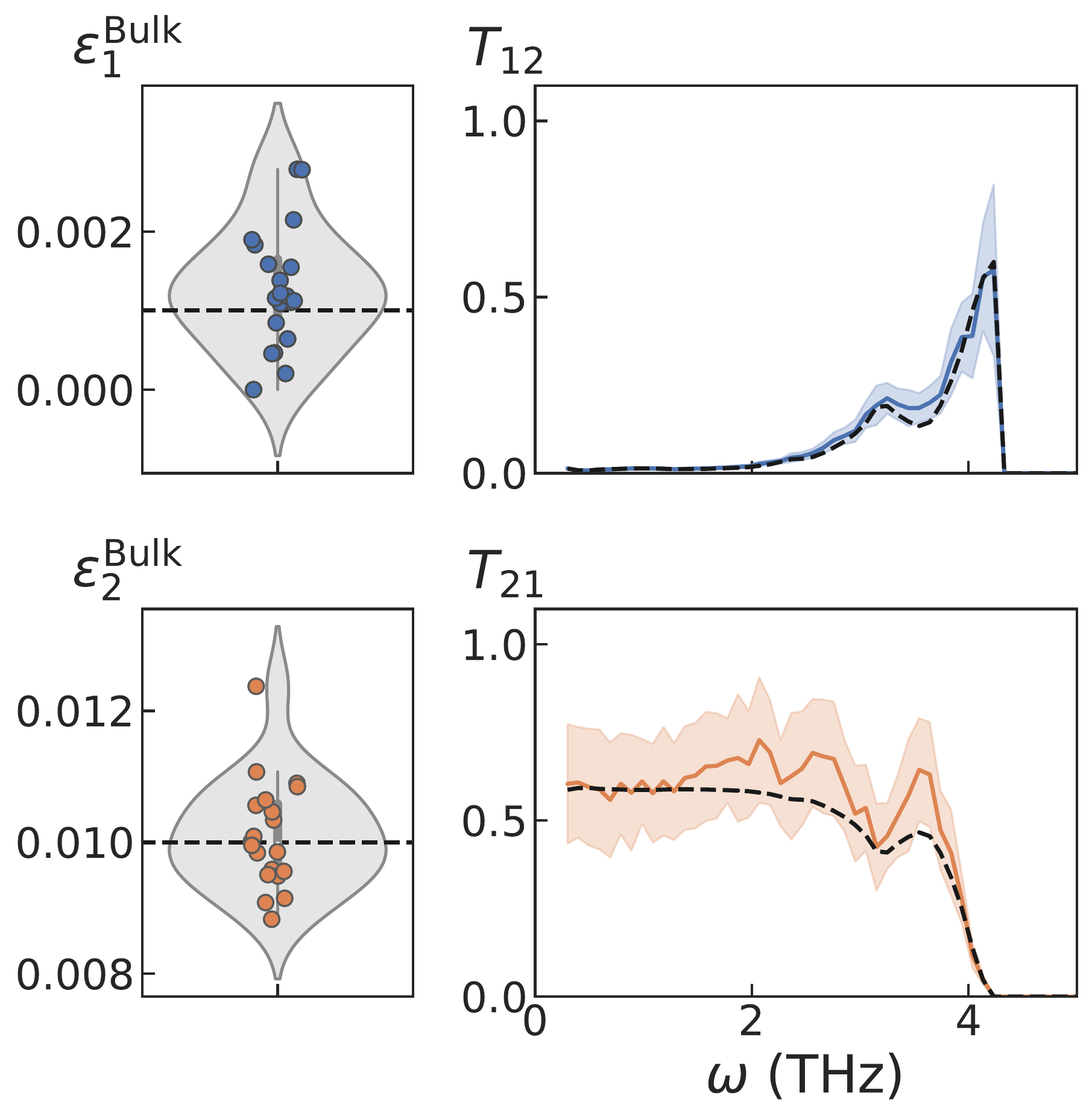}
    \caption{Comparisons between reconstructed and true bulk energy loss coefficients $\varepsilon_{1/2}^{\mathrm{Bulk}}$ and transmittance $T_{12/21}(\omega)$ from clean synthetic MSDs of a $5$\ nm Au/ $35$\ nm Si heterostructure. The synthetic MSDs have total time span of $30\,\mathrm{ps}$ with $\Delta t=1\,\mathrm{ps}$, with boundary energy loss disabled ($\varepsilon_{1/2}^{\mathrm{Bdry}}\equiv0$) and initial conditions listed in Table \ref{SI_tab:initial_conditions_Au_Si}. The \textbf{right} figure corresponds to synthetic MSDs that have same single initial condition and time points as experimental data. The statistics presented here come from 20 sets of randomly initialized parameters.}
    \label{fig:SI_AuSi_NumExp_EpsBulk_TraRef}
\end{figure}

\begin{figure}
    \centering
    \includegraphics[height=0.35\linewidth]{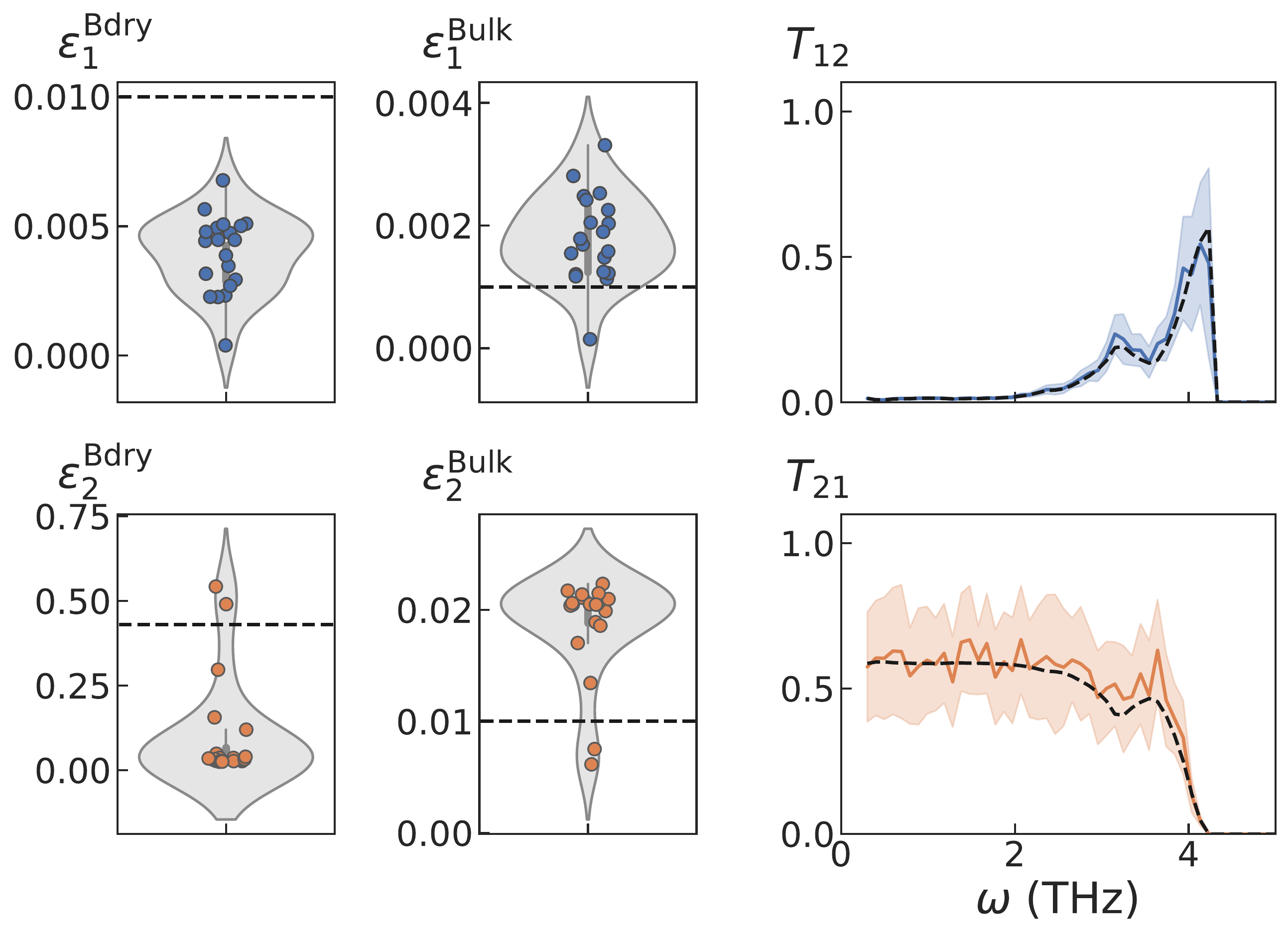}
    \hspace{5pt}
    \includegraphics[height=0.35\linewidth]{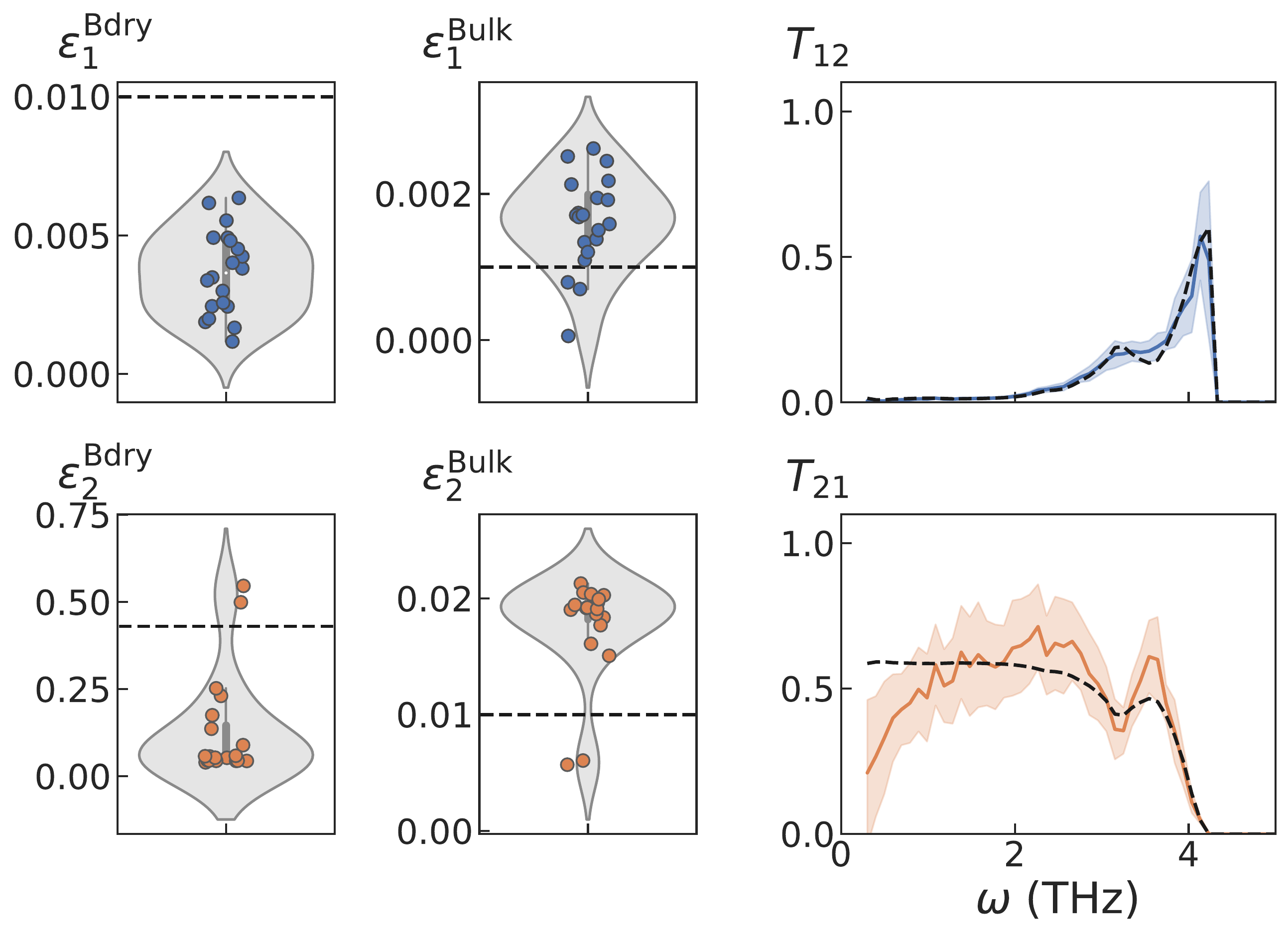}
    \caption{Comparisons between reconstructed and true boundary and bulk energy loss coefficients , $\varepsilon_{1/2}^{\mathrm{Bdry}}$ and $\varepsilon_{1/2}^{\mathrm{Bulk}}$, and transmittance $T_{12/21}(\omega)$ from clean synthetic MSDs of a $5$\ nm Au/ $35$\ nm Si heterostructure. The \textbf{left} figure corresponds to synthetic MSDs that have total time span of $30\,\mathrm{ps}$ with $\Delta t=1\,\mathrm{ps}$, with initial conditions listed in Table \ref{SI_tab:initial_conditions_Au_Si}. The \textbf{right} figure corresponds to synthetic MSDs that have same single initial condition and time points as experimental data. The statistics presented here come from 20 sets of randomly initialized parameters.}
    \label{fig:SI_AuSi_NumExp_EpsBdry_EpsBulk_TraRef}
\end{figure}

We carried out numerical experiments that reconstruct phonon properties from synthetic MSDs. Given only bulk energy loss, our framework can faithfully retrieve the corresponding energy loss coefficients $\varepsilon_{1/2}^{\mathrm{Bdry}}$ and the transmittance $T_{12/21}(\omega)$, as shown in Fig.\@ \ref{fig:SI_AuSi_NumExp_EpsBulk_TraRef}. In the presence of both boundary and bulk energy losses, we demonstrate in Fig.\@ \ref{fig:SI_AuSi_NumExp_EpsBdry_EpsBulk_TraRef} that the reconstructions of transmittance $T_{12/21}(\omega)$ is robust against inaccurate reconstructions of $\varepsilon_{1/2}^{\mathrm{Bdry}}$ and $\varepsilon_{1/2}^{\mathrm{Bulk}}$.

\begin{figure}
    \centering
    \includegraphics[width=0.4\linewidth]{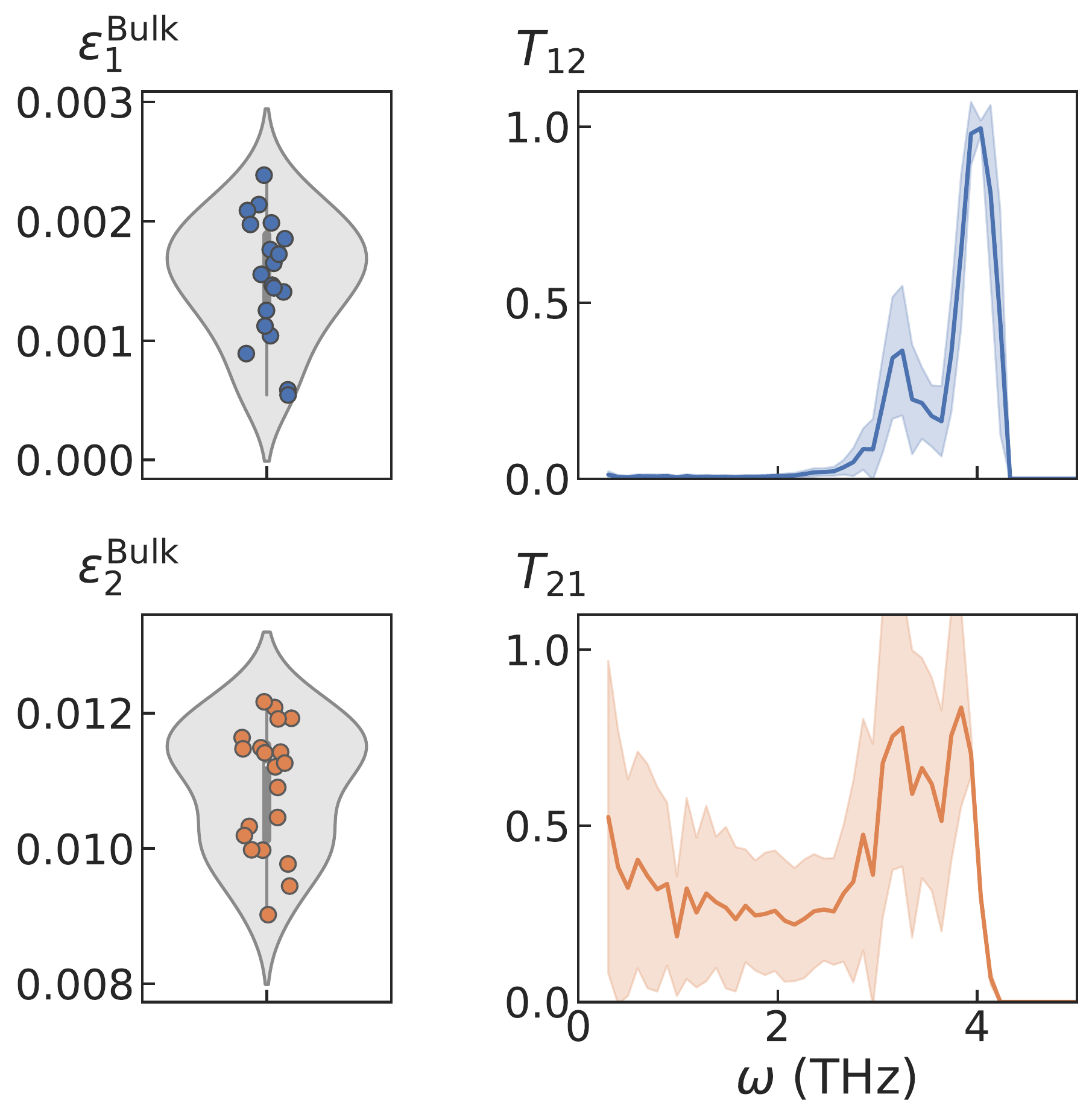}
    \caption{Reconstructions of bulk energy loss coefficients $\varepsilon_{1/2}^{\mathrm{Bulk}}$ and transmittance $T_{12/21}(\omega)$ from experimental MSDs. The statistics presented here come from 20 sets of randomly initialized parameters.}
    \label{fig:SI_AuSi_Exp_EpsBulk_TraRef}
\end{figure}

\begin{figure}
    \centering
    \includegraphics[width=0.6\linewidth]{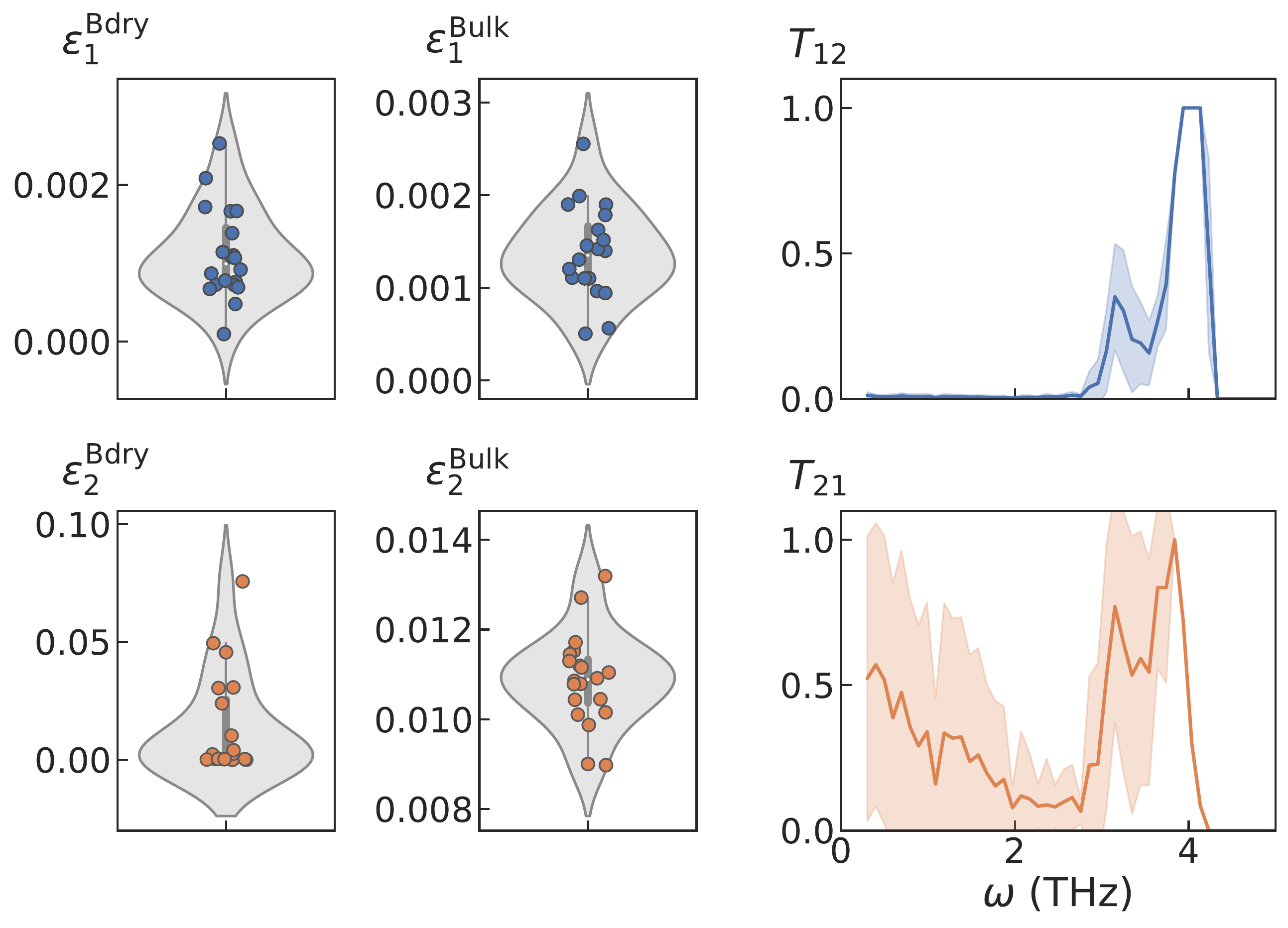}
    \caption{Reconstructions of boundary and bulk energy loss coefficients , $\varepsilon_{1/2}^{\mathrm{Bdry}}$ and $\varepsilon_{1/2}^{\mathrm{Bulk}}$, and transmittance $T_{12/21}(\omega)$ from experimental MSDs. The statistics presented here come from 20 sets of randomly initialized parameters.}
    \label{fig:SI_AuSi_Exp_EpsBdry_EpsBulk_TraRef}
\end{figure}

When investigating experimental MSDs, we can also turn on the fitting for $\varepsilon_{1/2}^{\mathrm{Bulk}}$. In particular, the Fig.\@ \ref{fig:SI_AuSi_Exp_EpsBulk_TraRef} shows reconstructed transmittance with $\varepsilon_{1/2}^{\mathrm{Bdry}}\equiv0$, which displays obvious similarities with the case where both $\varepsilon_{1/2}^{\mathrm{Bdry}}$ and $\varepsilon_{1/2}^{\mathrm{Bulk}}$ are fitted (Fig.\@ \ref{fig:SI_AuSi_Exp_EpsBdry_EpsBulk_TraRef}) and the case where only $\varepsilon_{1/2}^{\mathrm{Bdry}}$ is fitted (Fig.\@ 4(d) in main text).

